\documentclass{article} 
\usepackage[T1]{fontenc}
\usepackage{storm_preprint,times}

\usepackage{amsmath,amsfonts,bm}

\def\eqref#1{equation~\ref{#1}}

\def\1{\bm{1}}

\DeclareMathAlphabet{\mathsfit}{\encodingdefault}{\sfdefault}{m}{sl}
\SetMathAlphabet{\mathsfit}{bold}{\encodingdefault}{\sfdefault}{bx}{n}

\usepackage{hyperref}
\usepackage{url}
\usepackage{xurl}
\usepackage{graphicx}
\usepackage{wrapfig}
\usepackage{booktabs,tabularx,array}
\usepackage{pifont}
\usepackage{enumitem}

\newcolumntype{Y}{>{\centering\arraybackslash}X}

\title{STORM: A Lightweight Structure-Free Model for Binding Affinity Prediction}

\newcommand*\samethanks[1][\value{footnote}]{\footnotemark[#1]}

\author{%
  Ayush Pandit\textsuperscript{*,\textdagger} \\
  Stanford University \\
  \texttt{apandit@stanford.edu}
  \And
  Rohan V. Koodli\textsuperscript{*,\textdagger} \\
  Stanford University \\
  \texttt{koodli@stanford.edu}
  \And
  Wei Deng\samethanks[1] \\
  Stanford University \\
  \texttt{yaowei@stanford.edu}
  \And
  Roxana Daneshjou \\
  Stanford University \\
  \texttt{roxanad@stanford.edu}
  \And
  Ron O. Dror \\
  Stanford University \\
  \texttt{rondror@cs.stanford.edu}
}

\iclrfinalcopy
\begin{document}

\maketitle

\begingroup
\renewcommand{\thefootnote}{\fnsymbol{footnote}}
\footnotetext[1]{Equal contribution.}
\footnotetext[2]{Work conducted in part during an internship at Infrinia AI.}
\endgroup

\begin{abstract}
Accurately predicting a small molecule's binding affinity to a protein, a critical task in drug discovery, remains challenging. Physics-based approaches like free-energy perturbation (FEP) have very high computational costs and require expert setup. Recent machine learning--based cofolding models have reported accuracy approaching that of FEP---and at least one of them, Boltz-2, has already been widely adopted for affinity prediction---but they remain expensive for large computational screens and even more expensive for retraining on newly collected experimental data. To address these challenges, we introduce \textbf{STORM}: \textbf{S}imilarity \textbf{T}ransfer from \textbf{O}bserved \textbf{R}eference \textbf{M}olecules. \textsc{STORM} is a fast, simple machine learning model that ensembles many small model fusions of diverse protein and molecule representations for binding affinity prediction. Trained on the same data, \textsc{STORM} matches or exceeds the accuracy of Boltz-2 on multiple evaluation datasets, while delivering a speed improvement of over five orders of magnitude for both inference and training. Our results also suggest opportunities to substantially improve affinity prediction accuracy by integrating historical experimental affinity data with structural and biophysical information.

\end{abstract}

\section{Introduction}
Finding molecules that bind tightly to a target protein is a critical early step in drug discovery~\citep{paggi2024art}. Despite decades of research, however, accurately predicting binding affinities computationally remains an open challenge. As libraries of drug-like candidate molecules available to researchers grow into trillions of compounds~\citep{emolecules2022explore,enamine2024chemspace}, having fast and accurate methods for predicting binding affinities has become even more important. 

One major branch of research has focused on developing physics-based simulation methods for predicting binding affinities~\citep{wang2015fep}. Methods like free energy perturbation (FEP) can make predictions that correlate well with experimental measurements when provided a structure of the target protein bound to the small molecule. However, setting up FEP simulations requires careful expert calibration, high-quality structures, and significant computational resources---many hours of GPU time per prediction \citep{crivelli_decker2024aqfep, wang2015fep, heinzelmann2021automation}. Recently, machine learning methods such as Boltz-2~\citep{passaro2025boltz2} have reported performance approaching FEP on binding affinity prediction, at a far lower computational cost. These methods are cofolding foundation models, or machine learning architectures similar to AlphaFold3~\citep{abramson2024accurate,shen2026unlocking}, which was designed for biomolecular structure prediction. Boltz-2 pioneered the approach of extending these models to also predict binding affinities~\citep{passaro2025boltz2}. Subsequent studies have applied Boltz-2 to affinity prediction~\citep{fry2026zeroshot,amini2026affinity}, and several other cofolding models have incorporated affinity prediction using the same approach~\citep{shenoy2026nesso1,sandboxaq2026aqaffinity}.

Cofolding models are trained on large amounts of experimental biomolecular structure data and even larger amounts of biomolecular sequence data in the form of multiple sequence alignments (MSAs)~\citep{passaro2025boltz2, abramson2024accurate, openfold3-preview, wohlwend2024boltz1}. To add affinity prediction capabilities, they are also trained on datasets of publicly reported experimental binding affinities for compounds at many different protein targets. These methods are substantially faster than FEP, but they still require extensive computational resources to complete large virtual compound screens for drug discovery, limiting the number of compounds that can be screened~\citep{passaro2025boltz2, sandboxaq2026aqaffinity, lukauskis2025optimizing, singh2026navigating}. Because these methods need to be trained on very large datasets with many different types of data, refitting them or fine-tuning them on small amounts of additional experimental data as it becomes available can also be difficult~\citep{kumar2022finetuning, amini2026affinity, liao2026cocofold}. 

Most of the data used to train cofolding models--in particular,  experimental structures and MSAs--is critical for their structure-prediction capabilities. It is unclear, however, to what extent this data contributes to their affinity-prediction capabilities. Might a much simpler model that ignores three-dimensional structure, and that is trained only on experimental binding affinities, be able to achieve similar performance? Decades of research have shown that models trained only on affinity data at an individual target can achieve reasonable binding affinity predictions at that target by taking advantage of similar molecules being more likely to have similar binding affinities~\citep{cherkasov2014qsar,zhang2024topological}. Recent work has shown that these chemical trends in binding affinity be transferred between protein targets that share low sequence identity, suggesting that learning to transfer historical affinity relationships between protein targets could provide an alternative route to strong binding affinity prediction~\citep{mattsson2026identifying}. Focusing on this signal would also remove much of the training and modeling complexity associated with MSA and structure data. 

Inspired by these observations, we introduce \textbf{STORM}, or \textbf{S}imilarity \textbf{T}ransfer from \textbf{O}bserved \textbf{R}eference \textbf{M}olecules. \textsc{STORM} is a structure-free binding affinity predictor that focuses on learning relationships between small molecule affinities that can transfer between protein targets. We leverage the fact that different simple representations for small molecules, such as various chemical descriptors or different molecular fingerprint algorithms, can capture different activity relationships between molecules and provide complementary signals. We also apply this technique to build simple protein target representations, and then independently train and ensemble many small models on fusions of these target and molecule representations. We make \textbf{four main contributions}:

\begin{enumerate}[label=\textit{\textbf{(\arabic*)}}]
    \item We introduce a structure-free method for combining simple, complementary protein and small-molecule representations that can be trained on historical affinity measurements.
    \item Across four affinity benchmarks, \textsc{STORM} achieved higher Pearson correlations than the five retrained structure-free baselines. Compared with Boltz-2, \textsc{STORM} achieved a higher Pearson correlation on the ChEMBL35 time split and comparable correlations on OpenFE, FEP+4, and CASP16. \textsc{STORM}'s low inference and training costs support large virtual screens and model updates as new affinity measurements become available.
    \item We use matched ablations to quantify how different components of the STORM ensemble impact performance.
    \item We show that fine-tuning on small sets of affinity measurements improves Pearson correlation on held-out compounds for all three targets absent from the training set (Section~\ref{sec:target-finetuning}).
\end{enumerate}

In addition to these practical advances, our work has important longer-term implications, suggesting opportunities to substantially improve the accuracy of machine learning--based affinity prediction by combining the strengths of cofolding models with physics-based affinity predictors.

\section{Related Work} \label{sec:related work}

\paragraph{Cofolding affinity predictors.} 
Many machine learning methods have been developed that predict the affinity of a small molecule bound to a protein~\citep{ballester2010machine,stepniewska2018development,jimenez2018kdeep}. Other methods, such as those in \cite{lu2022tankbind}, \cite{lu2024dynamicbind}, and \cite{morehead2025flowdock}, predict both the affinity and the pose of the docked ligand in the pocket of a given input protein structure. However, Boltz-2 was the first model to integrate a base cofolding model with a small molecule binding affinity prediction module, leading to significantly improved accuracy in affinity prediction compared to previous machine learning methods~\citep{passaro2025boltz2}. This was an important advance, particularly given that Boltz-2 is faster than leading physics-based methods. Cofolding models use an AlphaFold 3--style trunk to integrate protein and small-molecule sequence information, MSA information, and, optionally, existing template structures to predict biomolecular complex structures~\citep{abramson2024accurate, passaro2025boltz2, openfold3-preview}. Boltz-2 pioneered the use of the latent representation of the Boltz-2 trunk in combination with their predicted protein--small molecule complex to train an affinity predictor module on historical binding affinity data. Subsequently, many others have adopted this core design, including AQAffinity on an OpenFold3 base~\citep{sandboxaq2026aqaffinity,openfold3-preview}. TerraBind uses a coarse-grained cofolding representation and predicts affinity from the trunk's latent representation~\citep{rossi2026terrabind}. The IsoDDE report also presents affinity results on several public benchmarks~\citep{isodde2026}. \textsc{STORM} instead uses simple representations and historical binding affinity measurements alone. 

\paragraph{Structure-free affinity predictors.}
Substantial prior work has also been done on developing structure-free affinity predictors. Many early methods, including chemogenomic models, focused on learning from datasets of binding affinity values across many targets using matrix factorization or graph learning based approaches~\citep{pahikkala2015toward,he2017simboost}. Methods such as DeepDTA and AttentionDTA use initial simple amino acid sequence encodings for proteins and SMILES string representations for molecules that are combined using attention, and then trained end-to-end for affinity prediction \citep{ozturk2018deepdta,zhao2022attentiondta}. GraphDTA is similar, but it represents molecules as graphs based on their chemical connectivity before fusing to the learned target representation \citep{nguyen2021graphdta}. Other approaches instead use large, pretrained encoders to generate embedding representations for proteins and molecules. BALM encodes protein sequences with ESM-2, a protein language model and small molecules with ChemBERTa-2, a small molecule language model, projects their representations together, and then trains end-to-end~\citep{gorantla2025learning,ahmad2022chemberta}; Structure-free DTA is similar but uses the original ChEMBERTa~\citep{bidgoli2026structure}. Finally, PSICHIC uses a hybrid representation in which small molecules are represented as graphs based on their chemical connectivity, but the protein's ESM-2 embeddings and predicted residue contacts are used to build a protein graph, which then becomes fused with the molecule representation~\citep{koh2024psichic}. Despite the many differences between these methods, each of them individually uses a single representation for the target protein and a single representation for the small molecule when making predictions, fed through a single model. \textsc{STORM} instead relies on making an ensemble of many different small models that use many different combinations of simple input representations. 

\section{Methods}

\textsc{STORM} combines 15 independently fitted multilayer perceptrons
with different protein and ligand inputs and a shared training objective.
Each model shares a core multi-layer perceptron (MLP) architecture, with a
variably sized first layer depending on the input representation.
Each component receives one or two votes. We selected the ensemble
membership, votes, and graph-smoothing rule based on holdouts of the training data
(Appendix~\ref{app:model_details}).

\subsection{Task}

Given a target protein $p$, a small molecule $m$, and an experimental endpoint $e$, \textsc{STORM} predicts a continuous experimental activity value $\hat{y}(p,m,e)$. We train on measured concentrations on a $\mathrm{pActivity}$ scale. This means larger values correspond to lower measured concentrations and tighter binding affinities. The endpoint identifies the type of measurement made in an experimental assay. We encode $K_i$, $K_d$, $\mathrm{IC}_{50}$, and $\mathrm{EC}_{50}$ separately; other endpoint types receive an all-zero endpoint vector. Each component model receives an endpoint encoding and a molecule representation. Most components also receive a target representation; target-blind components receive no target features.

\subsection{Input Representations}
The ensemble uses nine ligand representations and five protein-input
configurations built from accession-linked UniProtKB GO and InterPro
annotations, a deterministic 64-hot ID128 accession code, and a species
indicator~\citep{blum2025interpro,geneontology2026knowledgebase,uniprot2025}.
These input representations are independent of 3D structural data and
embeddings from pretrained protein language models or molecular foundation models.
For an unseen accession, the combined configuration can borrow the ID128
code of an MMseqs2 training hit. ID-only components hash the query
accession, and target-blind components receive only ligand and endpoint
inputs. Appendix~\ref{app:target-representations} describes the target
encodings. We compute ligand features with RDKit and
scikit-fingerprints~\citep{rdkit,adamczyk2024scikit}.

\begin{table}[t]
\centering
\small
\caption{Base target features used by \textsc{STORM}.}
\label{tab:storm-target-features}
\begin{tabularx}{\linewidth}{@{}lXr@{}}
\toprule
Feature & Description & Input size \\
\midrule
Target ID & Code derived from a hash of the target identifier. & 128 \\
GO & Gene Ontology annotations~\citep{geneontology2026knowledgebase,ashburner2000geneontology}. & 2,048 \\
InterPro & Protein-family and domain annotations~\citep{blum2025interpro}. & 2,048 \\
Species & Human, nonhuman, or unknown indicator. & 3 \\
\bottomrule
\end{tabularx}
\end{table}

\begin{table}[t]
\centering
\small
\caption{\textsc{STORM} Molecule representations.}
\label{tab:storm-ligand-features}
\begin{tabularx}{\linewidth}{@{}lXr@{}}
\toprule
Representation & Description & Input size \\
\midrule
Morgan count & Circular atom neighborhoods of radius 2~\citep{rogers2010extended}. & 2,048 \\
RDKit path count & Atom and bond paths of length 1--7~\citep{rdkit}. & 2,048 \\
Atom pair count & Atom-type pairs and their graph distances~\citep{carhart1985atom}. & 2,048 \\
Topological torsion count & Sequences of four connected atoms~\citep{nilakantan1987topological}. & 2,048 \\
Avalon binary & Avalon molecular fingerprint~\citep{rdkit}. & 2,048 \\
2D pharmacophore count & Pharmacophore feature pairs, triplets, and graph distances~\citep{skfp_pubchem}. & 2,048 \\
Klekota--Roth count & Predefined substructures~\citep{klekota2008chemical}. & 4,860 \\
PubChem count & PubChem chemical features~\citep{skfp_pubchem}. & 757 \\
RDKit/Mordred descriptors & Calculated 2D physicochemical and topological descriptors~\citep{rdkit,moriwaki2018mordred}. & 1,212 \\
\bottomrule
\end{tabularx}
\end{table}

\subsection{Model Architecture}

Every component model in \textsc{STORM} has a simple, early fusion architecture. The target representations, when present, are each projected to their referenced size and concatenated with the molecule representation in a shared MLP prediction head. The endpoint is encoded as a four-dimensional vector and projected into the first hidden layer. Table~\ref{tab:storm-architecture} summarizes the common prediction path. Here, $d_m$ is the size of the molecule representation and $d_p$ is the size of the encoded target representation; $d_p=0$ for target-blind components.  In total, across all 15 models, \textsc{STORM} has approximately 20 million parameters, or an average of 1.2 million parameters per component. Additional details are available in the Appendix. 

\begin{table}[t]
\centering
\small
\caption{Common architecture of each \textsc{STORM} component model.}
\label{tab:storm-architecture}
\begin{tabularx}{\linewidth}{@{}lccX@{}}
\toprule
Layer & Input size & Output size & Operation \\
\midrule
Input fusion & $d_m+d_p$ & 384 & Linear projection; add projected endpoint encoding \\
Hidden layer 1 & 384 & 384 & Layer normalization, SiLU, dropout ($0.20$) \\
Hidden layer 2 & 384 & 192 & Linear projection, layer normalization, SiLU, dropout ($0.10$) \\
Prediction & 192 & 1 & Linear projection to pActivity \\
\bottomrule
\end{tabularx}
\end{table}

\subsection{Loss Function}

We train each component using a combination of pointwise activity predictions and within-assay activity differences. Let $y_i$ and $\hat{y}_i$ denote the measured and predicted pActivity for example $i$. We use the Huber loss $H_{\delta}(r)$ with $\delta=1$. For a sampled pair $(i,j)$ from the same assay $a$, we normalize both the measured and predicted activity differences by
$s_a=\max(\sigma_a,0.5)$, where $\sigma_a$ is the standard deviation of measured activities in that training assay. The training objective for a batch is

\begin{equation}
\mathcal{L}
=
0.25\,\frac{1}{|B|}\sum_{i\in B}H_1(\hat{y}_i-y_i)
+
\frac{1}{|P_B|}\sum_{(i,j)\in P_B}
H_1\!\left(
\frac{(\hat{y}_i-\hat{y}_j)-(y_i-y_j)}{s_{a(i)}}
\right),
\end{equation}

where $B$ is the batch and $P_B$ is its set of sampled within-assay pairs. This helps us avoid high errors for small differences between values in assays, while still using a relative scale. We sample pairs only from assays with at least five training compounds. For an assay fragment containing $n$ examples in a batch, we sample up to $\min\{\binom{n}{2},\operatorname{round}(8\sqrt{n}),256\}$ pairs. The pointwise term learns absolute activity values, while the pairwise term directly trains the model to predict activity differences between molecules measured in the same assay.

\subsection{Graph Smoothing}
Graph smoothing is an optional inference step for a group of jointly
scored compounds. The benchmarks use one group per assay. Smoothing uses
only molecular fingerprints and the ensemble's predictions, without access
to measured assay activities. We compute
Tanimoto similarity $s_{ij}$ between the binary supports of the
Klekota--Roth count fingerprints~\citep{klekota2008chemical}. Let $N_i$
contain up to five other compounds in the group with the highest positive
similarity to compound $i$. Starting from the voted prediction
$\hat{y}_i^{(0)}$, we apply~\citep{taubin1995signal,vonluxburg2007tutorial,li2018deeper}

\begin{equation}
\hat{y}_i^{(t+1)}
=
(1-\lambda)\hat{y}_i^{(t)}
+
\lambda
\frac{\sum_{j\in N_i}s_{ij}\hat{y}_j^{(t)}}
{\sum_{j\in N_i}s_{ij}},
\qquad \lambda=0.10,
\end{equation}

for three passes, $t=0,1,2$. A singleton or a compound with no
positive-similarity neighbors retains its original ensemble score. Changing
the query group can change a compound's neighbors and score, and smoothing
can attenuate activity cliffs. For a score independent of the other
queried compounds, use the unsmoothed ensemble prediction. The paired
ablation is reported in Appendix~\ref{app:smoothing-ablation}.

\subsection{Target-Specific Finetuning}

\textsc{STORM} can be fine-tuned on newly measured affinities for new or existing targets. We begin with the base pretrained component models and finetune each independently on the new support affinities. We update the full model using a pointwise Huber loss and penalty for movement away from the weights' starting point \citep{pmlr-v80-li18a}. See the Appendix for more details. 

\section{Experiments}

\subsection{Training Data} \label{sec:training data}
We train \textsc{STORM} and all structure-free comparison methods on binding affinity measurements curated from the ChEMBL 34 \citep{chembl2024release34} release and the April 1, 2024 snapshot of BindingDB \citep{bindingdb2024april, liu2025bindingdb}. We then apply a data cleaning and standardization protocol adapted from the Boltz-2 model \citep{passaro2025boltz2} to remove low quality data, filter for specific experimentally measured endpoints, and standardize small molecules, protein targets, and affinity values.  Full details are available in the Appendix.

For OpenFE and FEP+4, we removed training proteins in clusters
containing benchmark targets at 90\% sequence identity under the
reported MMseqs2 protocol~\citep{steinegger2017mmseqs2}. For CASP16 and
ChEMBL 35 Time Split, we removed exact pairs of UniProt accession and standardized
ligand. See Appendices~\ref{app:data_preprocessing}
and~\ref{app:similarity} for the curation and overlap analyses.

In total, our training set is composed of 6,775 unique protein IDs, 934,000 unique molecules, 175,000 chemical series, and 1.7 million unique experimental binding affinity measurements.

We note that while our data sources were chosen to match Boltz-2 and enable a fair comparison, they are a \textit{subset} of the training data sources used to train Boltz-2's affinity module. In addition to these sources, Boltz-2 also uses data from PubChem small assays \citep{kim2023pubchem}, PubChem high-throughput screens \citep{kim2023pubchem}, CeMM fragment screens \citep{offensperger2024large}, MIDAS metabolite data \citep{hicks2023protein}, and synthetic decoys \citep{passaro2025boltz2}. These sources collectively represent millions of additional training examples and experimental data points, though many are binary values. Boltz-2's exact affinity training records have not been released, so we cannot complete an exact reproduction of the models' training data or confirm exact data overlap.

Importantly, we note that many of the choices that went into Boltz-2 data curation were designed specifically for the Boltz-2 model architecture, which jointly optimizes a continuous affinity value head and a binary binding likelihood head on a shared affinity module. Boltz-2 also filtered data based on the predicted quality of its generated protein-small molecule complex. These data curation rules do not transfer to structure-free binding affinity prediction models that output continuous values. For these reasons, we train \textsc{STORM} and all structure-free binding affinity prediction methods on only exact binding affinity measurements from ChEMBL 34 and BindingDB without matching all of the Boltz-2 filtering steps.

\subsection{Comparisons}

\subsubsection{Structure-based methods}

In Section \ref{sec:related work}, we referenced multiple cofolding models that have been trained for affinity prediction. However, many of them are ill-suited for benchmarking and comparison. 

Notably, IsoDDE reports evaluations on multiple benchmarks and has the highest reported performance on several \citep{isodde2026}. However, although its technical report specifies a training date cutoff, the actual composition of its training data is not disclosed. Consequently, IsoDDE's results cannot be compared with other methods: differences in performance may simply reflect differences in training data. 

While TerraBind has disclosed more information about its model architecture and training data \citep{rossi2026terrabind}, it is also a proprietary, closed-source model with limited benchmark reports, making comparisons to our full suite of benchmarks impossible. AQAffinity does share code, weights, and details on training data \citep{sandboxaq2026aqaffinity}. However, they train their model on later versions of ChEMBL and BindingDB, which overlap with our evaluation sets. Evaluating AQAffinity alongside Boltz-2 would require complete retraining, which is computationally infeasible. 

Because of these challenges, we use Boltz-2 as a representative model for structure-based methods. As an open-source model with weights freely available, Boltz-2 is also an important practical comparison, as the most widely deployed cofolding model with affinity prediction today. It has been benchmarked on all of our evaluations, and pioneered the affinity prediction approach used by all subsequent models today.

\subsubsection{Structure-free methods}

As referenced in Section \ref{sec:related work}, many different structure-free methods exist. For our comparison, we chose published models with the best reported performance. We compare \textsc{STORM} against \textbf{BALM}~\citep{gorantla2025learning}, \textbf{Structure-free DTA}~\citep{bidgoli2026structure}, \textbf{Attention DTA}~\citep{zhao2022attentiondta}, and \textbf{PSICHIC}~\citep{koh2024psichic}. As an additional baseline, we also compare to the molecular fingerprint and descriptor method in \cite{mattsson2026identifying}, referred to as \textbf{Mattsson-Walters}. This method uses a molecular fingerprint, descriptors, and a one-hot protein representation, which are fed into an MLP to make binding affinity predictions. We retrained all five methods on the shared curated corpus. See the Appendix for additional details.






\begin{figure*}[hbt!]
    \centering
    \includegraphics[width=1.0\linewidth]{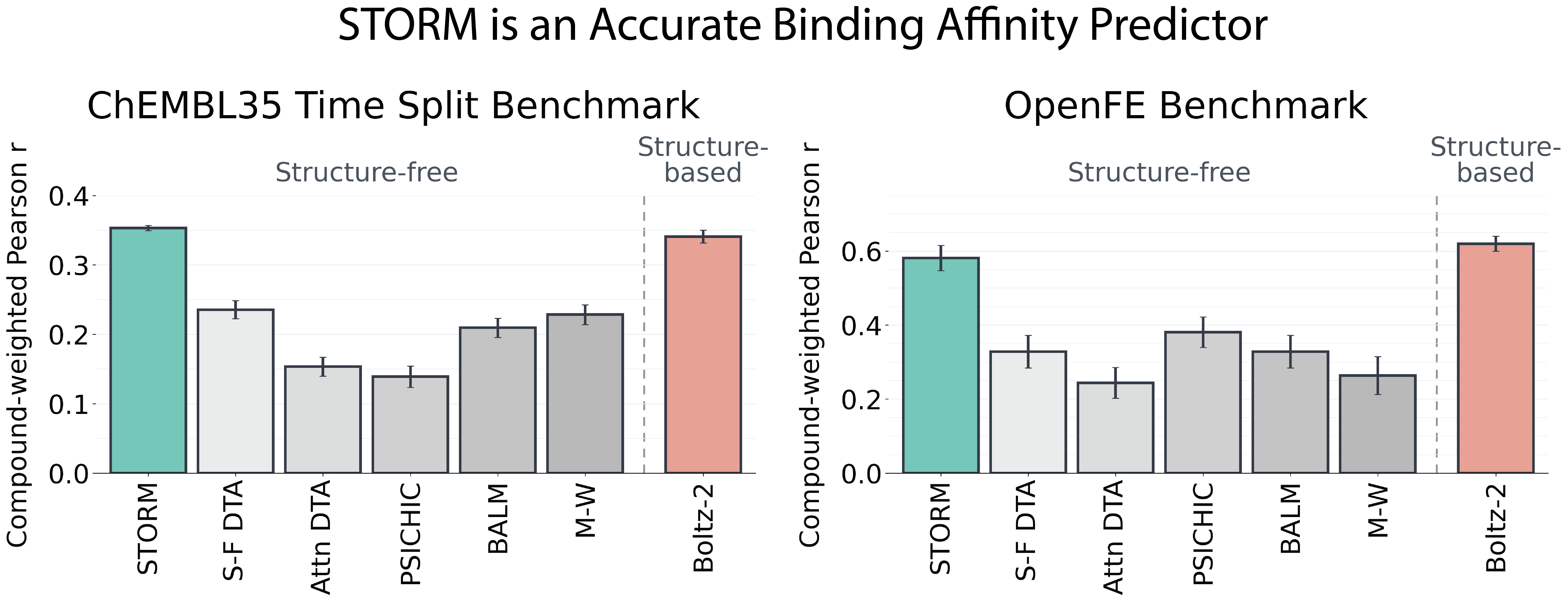}
    \caption{Historical full-benchmark compound-weighted within-assay Pearson
    $r$. S-F DTA denotes Structure-free DTA, Attn DTA denotes AttentionDTA,
    and M-W denotes Mattsson--Walters. Error bars for the locally evaluated
    models show bootstrap standard errors. Published Boltz-2 error bars were
    estimated from the source figures for descriptive comparison.
    Table~\ref{tab:model-performance} gives the matched evaluation used
    for paired statistical claims.}
    \label{fig:benchmark_results}
\end{figure*}

\subsection{Binding Affinity Benchmark Results}

We evaluated four benchmarks used by Boltz-2 and IsoDDE. OpenFE contains
873 measurements across 58 assays and 39 targets, including kinases and
other enzymes and excluding membrane proteins such as
GPCRs~\citep{ross2023fepbenchmark,gowers2023openfe}. FEP+4 is a subset of
the FEP+ benchmark with 87 measurements across four assays and four kinase
targets (CDK2, TYK2, JNK1, and
p38\ensuremath{\alpha})~\citep{ross2023fepbenchmark,hahn2022best}. Both benchmarks are
exact-target holdouts under our training exclusion protocol. CASP16 is a
retrospective evaluation of 140 measurements across two training-seen
targets, human chymase (17 measurements) and human autotaxin (123
measurements)~\citep{gilson2026casp16}. ChEMBL35
Time Split consists of newly deposited assays in ChEMBL 35, with
15,304 measurements across 449 assays and 278 distinct protein
IDs~\citep{chembl35,isodde2026}. For the matched comparison, we scored
every model on the same 16,245 measurements after excluding rows without
Boltz-2 predictions (Table~\ref{tab:matched-benchmark-coverage}) 
(Appendix~\ref{app:similarity}).

\begin{table}[t]
\centering
\small
\setlength{\tabcolsep}{2pt}
\caption{Compound-weighted Pearson $r$ on benchmarks. Values are estimates
$\pm$ one within-assay bootstrap standard error. See
Appendix~\ref{app:matched-benchmark-evaluation} for more details.}
\label{tab:model-performance}
\vspace{5pt}
\begin{tabular}{@{}lcccc@{}}
\toprule
Model & ChEMBL35 & OpenFE & FEP+4 & CASP16 \\
\midrule
STORM & $\mathbf{0.355\pm0.007}$ & $\mathbf{0.576\pm0.024}$ & $\mathbf{0.705\pm0.053}$ & $\mathbf{0.671\pm0.044}$ \\
Boltz-2 (local, corrected) & $0.303\pm0.007$ & $0.570\pm0.024$ & $0.675\pm0.062$ & $0.640\pm0.052$ \\
Structure-free DTA & $0.238\pm0.008$ & $0.328\pm0.030$ & $0.441\pm0.092$ & $0.380\pm0.075$ \\
AttentionDTA & $0.157\pm0.008$ & $0.244\pm0.031$ & $0.062\pm0.089$ & $0.580\pm0.049$ \\
PSICHIC & $0.142\pm0.008$ & $0.381\pm0.029$ & $0.518\pm0.072$ & $0.565\pm0.054$ \\
BALM & $0.210\pm0.008$ & $0.328\pm0.031$ & $0.481\pm0.086$ & $0.387\pm0.066$ \\
Mattsson--Walters & $0.230\pm0.008$ & $0.264\pm0.032$ & $0.638\pm0.073$ & $0.252\pm0.085$ \\
\midrule
Boltz-2 (published reference) & $\approx 0.341$ & $0.620$ & $0.660$ & $0.650$ \\
\bottomrule
\end{tabular}
\end{table}

\paragraph{Within-assay correlation.}
Compound-weighted Pearson correlation measures relative prioritization
within assays and is invariant to additive offsets within an assay
\citep{ross2023fepbenchmark}. Figure~\ref{fig:benchmark_results} shows
the historical full-benchmark comparison, including the three-fit STORM
summary. The figure emphasizes ChEMBL35 Time Split and OpenFE because they
cover more protein targets (278 and 39, respectively) than CASP16 and FEP+4
(two and four, respectively). In the matched comparison (Table~\ref{tab:model-performance}),
STORM had higher Pearson point estimates than the five retrained
structure-free baselines in all 20 comparisons, with 18 significant
after Holm correction.

\paragraph{Direct comparison with Boltz-2.}
We ran Boltz-2 v2.2.1 with ColabFold MSAs and applied the optional
molecular-weight correction to the saved affinity predictions. The
correction increased Boltz-2's Pearson correlation on all four
benchmarks (Table~\ref{tab:boltz-mw-sensitivity}). We scored the
seed-2023 STORM and Boltz-2 predictions on the same measurements, using
the same assay groupings and compound-count weights. This follows the
weighting described in the Boltz-2 paper~\citep{passaro2025boltz2}.
For ChEMBL35, we reconstructed the published reference of approximately
0.341 from the per-class Boltz-2 scores in IsoDDE Figure~8. We report
the published values as descriptive references and base statistical
comparisons on the local paired evaluation.

On ChEMBL35 Time Split, STORM scored 0.3549 and local Boltz-2 scored
0.3028, a difference of $+0.0521$ in Pearson $r$ (unadjusted 95\% paired
bootstrap-$t$ CI $[+0.0359,+0.0691]$, Holm-adjusted $p=0.00288$).
STORM and Boltz-2 had comparable Pearson correlations on OpenFE, FEP+4,
and CASP16. For full results, see Appendix~\ref{app:matched-benchmark-evaluation}.

\paragraph{Ensemble controls.}
The diverse ensemble scored 0.5814 on OpenFE and 0.3529 on ChEMBL35,
compared with 0.5055 and 0.2562 for 15 pooled fits of the strongest
standalone target-blind RDKit path component. In the matched three-fit
controls, protein inputs improved ChEMBL35 by 0.0619 and OpenFE by 0.0464
after adjustment. Ligand representation diversity improved ChEMBL35 by
0.0411 after adjustment, while the OpenFE contrast was inconclusive.
The control protocols and all four benchmark results appear in
Appendix~\ref{app:matched-ablation}.

\subsection{Inference Throughput} \label{sec:fast inference}

\paragraph{Setup} We timed inference on one NVIDIA L4 GPU with 24~GB
of memory. We compared the largest feasible batch and the largest feasible
power-of-two batch. For each batch configuration, we repeated inference
runs and divided each run's elapsed time by the number of ligands
processed. We report the lower median elapsed time per ligand.
The Boltz-2 timing used no MSAs. The separate accuracy evaluation used
ColabFold MSAs (Appendix~\ref{app:matched-benchmark-evaluation}).

\begin{figure*}[hbt!]
    \centering
    \includegraphics[width=1.0\linewidth]{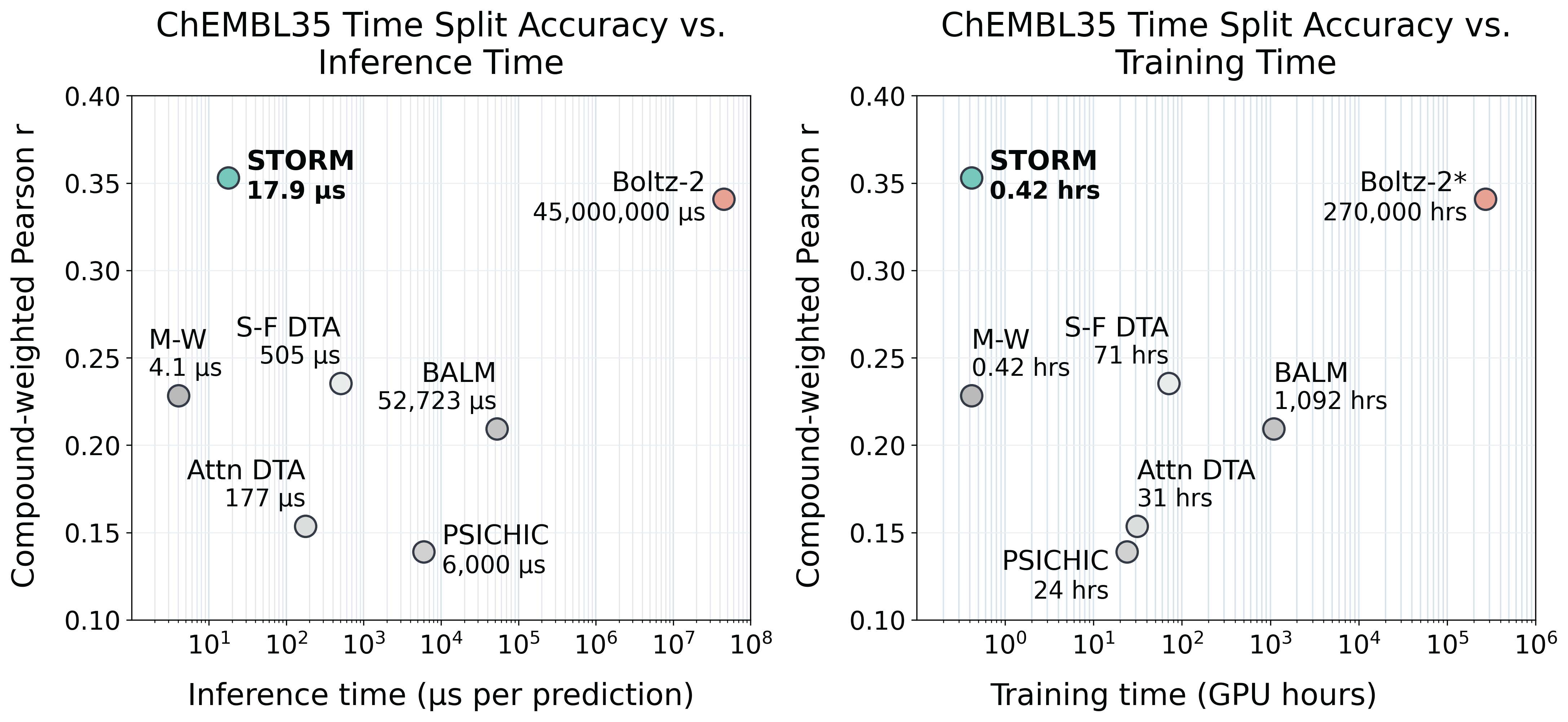}
    \caption{Historical ChEMBL35 within-assay Pearson $r$ versus measured
    inference time per ligand (left) and training compute (right).
    The Boltz-2 inference timing uses no MSAs. The Boltz-2 training cost
    is an L4-equivalent estimate.}
    \label{fig:speed}
\end{figure*}

\paragraph{Results} STORM scored 17.9 microseconds per prediction and
was second in throughput to Mattsson--Walters. Boltz-2 took approximately
45 seconds per ligand in the no-MSA timing configuration.

\subsection{Training Compute}

We compared the training wall times of the structure-free methods on
the NVIDIA L4 and used the fastest measured batch configuration for
each method. STORM trained in under half an hour. The approximately
270,000 L4-equivalent GPU hours for Boltz-2 are an estimate based on
the scaling assumptions in Appendix~\ref{app:boltz-compute}. The
separate laptop benchmark projected 45-epoch wall times of
2~h~04~min~36~s on CPU and 1~h~20~min~23~s on PyTorch MPS from
short training runs (Appendix~\ref{app:laptop-training}).

\subsection{Target-Specific Fine-Tuning Results}
\label{sec:target-finetuning}

\setlength{\intextsep}{1pt}
\begin{figure}
  \centering
  \includegraphics[width=300pt]{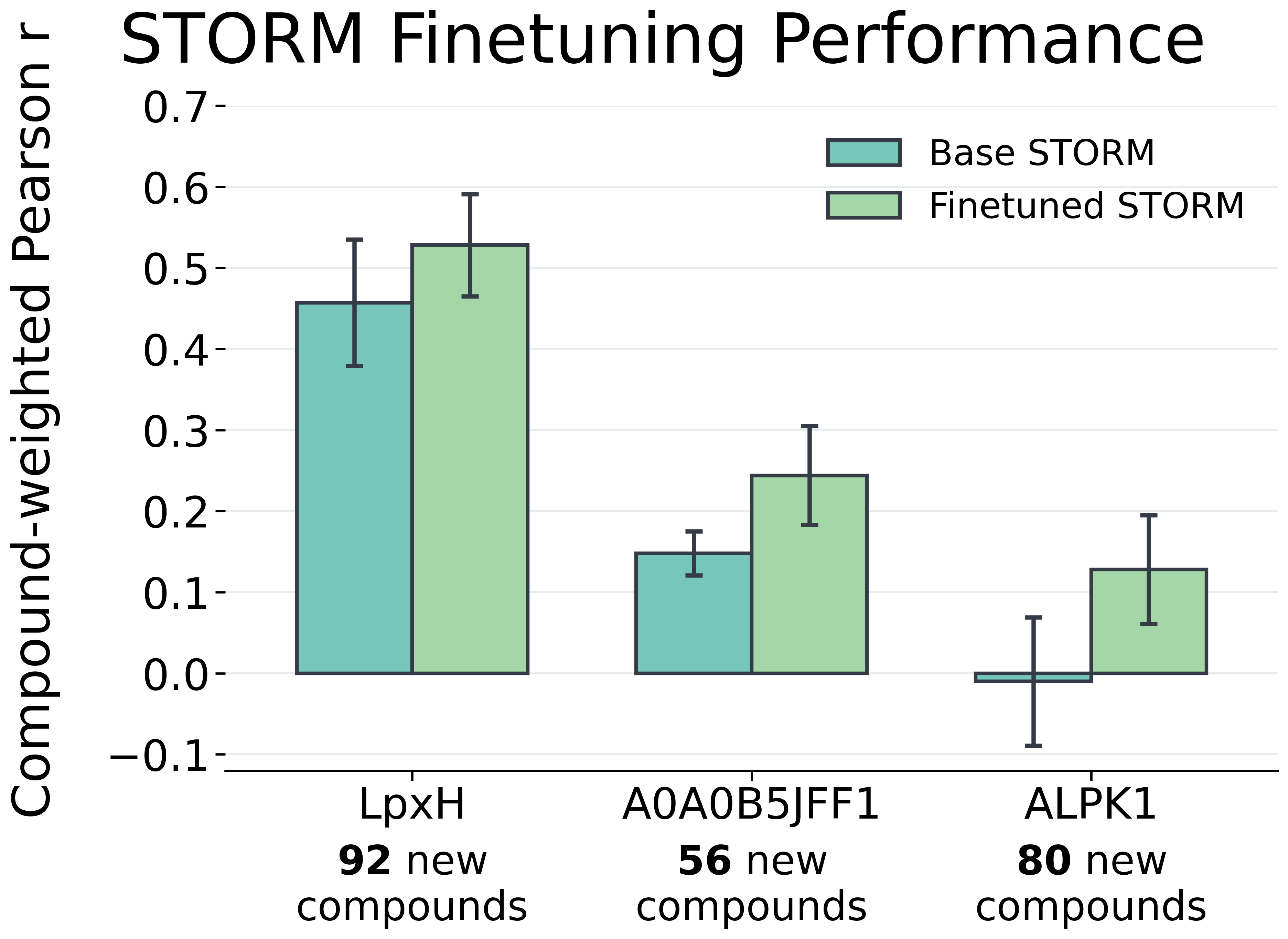}
  \caption{\textsc{STORM} finetuning results on three new targets. Error bars denote standard deviation.}
  \label{fig:finetuning-targets}
\end{figure}


\paragraph{Setup} Drug discovery campaigns often have limited affinity
measurements for a target, and the cost of collecting more motivates
fine-tuning on small datasets. We fine-tuned STORM separately on measurements for
LpxH, A0A0B5JFF1, and ALPK1. The targets and their selected ligands
were absent from the training corpus. We used the support and test
splits described in Appendix~\ref{app:finetuning-setup} to compare
the adapted ensembles with their corresponding base ensembles.

\paragraph{Results} Fine-tuning improved Pearson correlation on the
test compounds for all three targets (Figure~\ref{fig:finetuning-targets}).
For ALPK1, fine-tuning on 80 compounds substantially improved Pearson
correlation from the base ensemble's near-zero value.

\section{Conclusion}

We present \textsc{STORM} as a practical choice for machine learning--based binding affinity prediction. \textsc{STORM} achieves performance exceeding or comparable to the state-of-the-art machine learning methods at a fraction of the computational cost for training and inference. \textsc{STORM} achieves this without using embeddings generated by pretrained protein language models or molecular foundation models, and with a simple model architecture that is easy for researchers to adapt or modify. \textsc{STORM}'s ability to be retrained or easily finetuned on new experimental data--a critical use case in drug discovery--with ordinary consumer hardware provides a compelling advantage over other methods and makes it highly accessible for researchers everywhere. 

\textsc{STORM} also highlights an important gap between current machine learning methods and traditional physics-based simulation methods such as FEP. Unlike the machine learning methods discussed in this paper, physics-based methods are not trained on a large affinity dataset; they rely entirely on the provided structure of the protein target and biophysical principles to produce accurate predictions of binding affinity. On the contrary, \textsc{STORM} receives no signal from biomolecular structure or biophysical principles. Our results suggest opportunities for substantially improving cofolding methods: structural and biophysical information could complement historical affinity transfer, particularly in cases where reference information is sparse or similar molecules have sharply differing affinity measurements.

\clearpage
\bibliography{storm_references}
\bibliographystyle{storm_references}

\clearpage
\appendix
\setlength{\belowcaptionskip}{5pt}
\section{Training data curation}
\label{app:data_preprocessing}

We built one dataset of experimentally measured, equality-qualified
binding affinities from ChEMBL 34 and the 1 April 2024 BindingDB
snapshot. \textsc{STORM} and all structure-free comparison methods
were trained using this dataset. No model in this comparison was
trained on inequality-qualified measurements. We did not apply a
publication-date cutoff. ChEMBL35 Time Split is the benchmark name
used by IsoDDE~\citep{isodde2026}. ChEMBL 34 and BindingDB are a
subset of the sources used to train the Boltz-2 affinity module, which also
used data from PubChem, CeMM and MIDAS, as well as synthetic decoys
\citep{passaro2025boltz2}. The exact Boltz-2 affinity training records
are unavailable. The steps below distinguish curation shared
by the comparison methods from processing used for the canonical
\textsc{STORM} fit.

\subsection{General-purpose curation}

\begin{enumerate}[leftmargin=*]
    \item \textbf{Select measurements and assays.}
    We retained equality-qualified, positive concentrations in
    convertible units for $K_i$, $K_d$, IC$_{50}$, and EC$_{50}$
    from both sources, and AC$_{50}$ and XC$_{50}$ from ChEMBL.
    ChEMBL records required a single-protein target, one UniProt
    mapping, assay confidence score 9, direct target assignment,
    and a binding or functional assay. We excluded records flagged
    as potential duplicates or carrying validity annotations other
    than ``Manually validated.'' ChEMBL mutant and variant assays
    were retained and flagged. BindingDB records required a
    single-chain target and a documented assay mapping.

    \item \textbf{Resolve protein targets.}
    We assigned canonical UniProt sequences to retained targets
    and excluded records without an unambiguous mapping.
    BindingDB targets were resolved through trusted UniProt
    accessions or sequence matches to ChEMBL 34 and unambiguous
    BindingDB Swiss-Prot anchors. Sequence-based mappings required
    at least 90\% identity and 90\% coverage of both sequences.
    Canonical sequences were used as model inputs, including for
    records flagged as mutant or variant assays.

    \item \textbf{Standardize ligands and values.}
    Ligands underwent RDKit cleanup, largest-fragment selection
    with an organic-fragment preference, functional-group
    normalization, metal disconnection, and reionization, followed
    by canonical isomeric SMILES generation. We excluded
    unparseable structures, molecules with more than 50 heavy
    atoms, and molecules with exact molecular weight above
    1,000~Da. Concentrations were converted to
    $y=\log_{10}(c/1\,\mu\mathrm{M})$, with endpoint identities
    retained. We did not apply a PAINS filter, a model-derived
    structural-confidence filter, or a minimum assay-size threshold.

    \item \textbf{Remove evaluation overlap.}
    Using MMseqs2, we removed training proteins in clusters
    containing OpenFE or FEP+4 targets at 90\% sequence identity
    (\texttt{cov-mode=0}, \texttt{c=0.01}). Across both sources,
    we also removed records sharing a UniProt target and
    standardized ligand with CASP16 or IsoDDE evaluation records.
    Both CASP16 targets occur in the retained training set, as do some
    ChEMBL35 targets. The OpenFE and FEP+4 target holdouts retain some
    homologs below the exclusion threshold. Ligands and close analogs can
    recur at other targets. Appendix~\ref{app:similarity} quantifies this
    overlap relative to the retained training corpus.

    \item \textbf{Define series and remove duplicate assays.}
    We grouped measurements by source assay, target, and endpoint.
    BindingDB series used Entry DOI, canonical target, and endpoint,
    with Article DOI and then patent identifiers as fallbacks.
    Within- and cross-source assays were considered duplicates
    only when their complete standardized compound sets matched
    and paired concentrations agreed within 5\%. ChEMBL was
    preferred for cross-source duplicates. Partially overlapping
    assays and assays with differing measurements remained
    separate. We preserved source record and assay identifiers
    for traceability.
\end{enumerate}

\subsection{Canonical \textsc{STORM} processing}

\begin{enumerate}[leftmargin=*,resume]
    \item \textbf{Consolidate repeated assay--compound measurements.}
    For the canonical \textsc{STORM} fit, repeated measurements
    of the same standardized molecule within a series were
    averaged when their sample standard deviation was at most
    0.5 in $\log_{10}(\mu\mathrm{M})$ units. Groups exceeding this
    threshold were removed.

    \item \textbf{Construct the regression objective.}
    The canonical model combines pointwise Huber loss on exact
    affinity values with pairwise Huber loss on differences
    between compounds in the same assay. Assays with fewer than
    five compounds can contribute to the pointwise term but do
    not supply sampled pairs. No bounded-value or binary
    classification loss is used.
\end{enumerate}

\section{Protein Target and Molecule Representations}
\label{app:target-representations}

\paragraph{GO and InterPro representations.}
We use GO and InterPro annotations associated with the query's canonical
UniProtKB accession~\citep{uniprot2025,geneontology2026knowledgebase,blum2025interpro}.
We keep terms that occur in training targets and collapse terms with the same
training-target incidence into one feature. We then additively hash GO and
InterPro into separate 2,048-dimensional vectors. An annotation block with no
available retained terms is all zeros. A future implementation could obtain
annotations for a novel raw sequence with InterProScan.

\paragraph{Target ID representation.}
ID128 is a deterministic binary hash of the UniProt accession with 64 active
entries in 128 dimensions. The combined configuration can use a training
target's ID128 code for an unseen accession. The MMseqs2 donor search ranks
hits by decreasing bitscore, percent identity, minimum bidirectional coverage,
and alignment length, with target ID breaking remaining ties
\citep{steinegger2017mmseqs2}. The combined configuration uses the query's
own available annotations and species indicator. The ID-only configuration
hashes the query accession directly. The similarity analysis in
Appendix~\ref{app:similarity} summarizes aligned identity at the
highest-bitscore hit.

\paragraph{Target configurations.}
Our five target configurations are all four base features together; GO alone; InterPro alone; target ID alone; and a target-blind configuration with no target features. The species indicator is included only in the combined configuration.

\paragraph{Fingerprint Representations} Our fingerprint representations consist of some count-based fingerprints, which have positive integer feature counts, and binary fingerprints that indicate the presence or absence of a feature. The 2D pharmacophore fingerprint is folded into 2,048 additive count bins. Our merged RDKit/Mordred descriptors are formed from 411 selected RDKit and 934 selected Mordred descriptors. These are a subset of the total features calculated by the RDKit package and the Mordred package that were chosen for having less than 1\% of values missing, and for columns that had correlations of less than 98\%. Then, we completed a second de-duplication pass merging the RDKit features and Mordred features at the same threshold, leaving the final 1,212 descriptors we used.

\section{Model Details}
\label{app:model_details}

\paragraph{Canonical model inputs.}
For the canonical model, inequality-qualified measurements were excluded
from fitting. Repeated assay--compound measurements were averaged in
logarithmic units when their sample standard deviation was at most
0.5; groups exceeding this threshold were removed. This yielded
2,216,239 training rows across 174,946 series. 

\begin{table}[htbp]
\centering
\small
\label{tab:MLP reps}
\begin{tabularx}{\linewidth}{@{}
    >{\raggedright\arraybackslash}p{0.34\linewidth}
    >{\raggedright\arraybackslash}X
    r
@{}}
\toprule
Target representation & Ligand input & Votes \\
\midrule
Canonical (ID128 + GO + InterPro + species) & Atom pair count & 1 \\
Canonical & Morgan count & 1 \\
Canonical & 2D pharmacophore count & 1 \\
Canonical & RDKit/Mordred graph-collapsed 1,212 descriptors & 2 \\
Canonical & Topological torsion count & 1 \\
Target blind & Atom pair count & 1 \\
Target blind & Avalon binary & 1 \\
Target blind & Morgan count & 1 \\
Target blind & RDKit path count & 2 \\
GO only & 2D pharmacophore count & 1 \\
GO only & RDKit path count & 2 \\
InterPro only & Klekota-Roth count & 1 \\
InterPro only & PubChem count & 1 \\
InterPro only & RDKit/Mordred graph-collapsed 1,212 descriptors & 1 \\
ID128 only & Atom pair count & 1 \\
\bottomrule
\end{tabularx}
\end{table}

\paragraph{Weighted Ensemble.}
\textsc{STORM} is a weighted ensemble of 15 independently trained
components, each with one or two votes. We selected the component subset,
vote pattern, and graph-smoothing rule retrospectively on a random subset of the training data that was initially held out.  All components use a shared checkpoint epoch of 45 for seeds 2023, 2024, and 2025. The bootstrap intervals condition on the selected ensemble.

The component models are trained independently. For model $k$, let $\hat{y}_{ik}$ be its prediction for example $i$ and let $v_k\in\{1,2\}$ be its number of votes. The ensemble prediction is
\begin{equation}
\hat{y}_i^{(0)}
=
\frac{\sum_{k=1}^{15}v_k\hat{y}_{ik}}
{\sum_{k=1}^{15}v_k},
\qquad
\sum_{k=1}^{15}v_k=18.
\end{equation}
A two-vote component contributes twice to this average; it is still a single trained model.

\section{Ablations}

\paragraph{Contribution of ensemble components.}
To examine how the components contribute to \textsc{STORM}, we start with the 15-model, 18-vote ensemble and remove one model at a time. At each step, we remove the model whose deletion gives the highest mean of OpenFE and ChEMBL35 Time Split performance among the remaining models. The surviving models retain their original vote weights, and we apply the same graph-smoothing rule after each removal. Table~\ref{tab:storm-component-ablation} shows that performance changes little when the first few components are removed, but falls as the ensemble becomes smaller. With five models remaining, OpenFE Pearson $r$ decreases from $0.5814$ to $0.5639$, and ChEMBL35 Pearson $r$ decreases from $0.3529$ to $0.3245$. The removal order is greedy and depends on the models already removed, so it should not be interpreted as an independent measure of each component's importance. These ablations are used to demonstrate the performance contributions of different components of the STORM ensemble. They were not used during the initial selection of the model or model components.

\begin{table}[t]
\centering
\small
\caption{Greedy component-removal ablation of the selected \textsc{STORM} ensemble. At each step, the component whose removal maximizes the mean of OpenFE and ChEMBL35 Time Split performance is removed. The remaining predictions retain their original votes and receive the fixed graph-smoothing step. Values are compound-weighted within-assay Pearson $r$, reported as mean $\pm$ sample standard deviation across three independently trained seeds.}
\label{tab:storm-component-ablation}
\begin{tabularx}{\linewidth}{@{}rrXcc@{}}
\toprule
Models left & Votes left & Component removed & OpenFE & ChEMBL35 \\
\midrule
15 & 18 & Full ensemble & $0.5814 \pm 0.0089$ & $0.3529 \pm 0.0040$ \\
14 & 17 & Target-blind atom pair & $0.5815 \pm 0.0086$ & $0.3540 \pm 0.0034$ \\
13 & 16 & Canonical 2D pharmacophore & $0.5835 \pm 0.0091$ & $0.3517 \pm 0.0049$ \\
12 & 15 & ID-only atom pair & $0.5816 \pm 0.0077$ & $0.3519 \pm 0.0033$ \\
11 & 14 & InterPro-only Klekota--Roth & $0.5827 \pm 0.0106$ & $0.3477 \pm 0.0030$ \\
10 & 13 & InterPro-only merged descriptors & $0.5798 \pm 0.0103$ & $0.3464 \pm 0.0029$ \\
9  & 12 & Target-blind Avalon & $0.5750 \pm 0.0104$ & $0.3469 \pm 0.0026$ \\
8  & 11 & Canonical Morgan & $0.5772 \pm 0.0068$ & $0.3396 \pm 0.0043$ \\
7  & 10 & GO-only 2D pharmacophore & $0.5756 \pm 0.0087$ & $0.3348 \pm 0.0040$ \\
6  & 9  & InterPro-only PubChem & $0.5708 \pm 0.0115$ & $0.3293 \pm 0.0048$ \\
5  & 8  & Target-blind Morgan & $0.5639 \pm 0.0103$ & $0.3245 \pm 0.0067$ \\
4  & 7  & Canonical atom pair & $0.5565 \pm 0.0136$ & $0.3149 \pm 0.0056$ \\
3  & 6  & Canonical topological torsion & $0.5432 \pm 0.0137$ & $0.3025 \pm 0.0086$ \\
2  & 4  & GO-only RDKit path & $0.5293 \pm 0.0125$ & $0.2824 \pm 0.0117$ \\
1  & 2  & Canonical merged descriptors & $0.4949 \pm 0.0057$ & $0.2149 \pm 0.0055$ \\
\bottomrule
\end{tabularx}
\end{table}

\paragraph{Diverse representations versus repeated fits of one model.}
We ask whether the gain from combining different representations could instead be obtained by training more instances of a single strong component. Among the 15 selected components, the target-blind RDKit path model has the highest mean standalone performance across OpenFE and ChEMBL35 Time Split. We train 15 independent instances of this component with seeds 2023--2037 at epoch 45, equally average their predictions into one pooled score vector, and apply the same graph-smoothing rule used for \textsc{STORM}. As shown in Table~\ref{tab:storm-diversity-ablation}, this ensemble reaches $0.5055$ on OpenFE and $0.2562$ on ChEMBL35, compared with $0.5814$ and $0.3529$ for the diverse 15-component ensemble. The result suggests that combining complementary molecule and target representations contributes more than repeatedly fitting this single component.

\begin{table}[t]
\centering
\small
\caption{A diverse 15-component ensemble compared with 15 independent fits of the strongest single component, target-blind RDKit path. All predictions receive the same graph-smoothing step. The repeated-component column is one pooled prediction from 15 independent fits; the \textsc{STORM} column reports mean $\pm$ sample standard deviation across three independently trained ensembles. Values are compound-weighted within-assay Pearson $r$.}
\label{tab:storm-diversity-ablation}
\begin{tabular}{@{}lcc@{}}
\toprule
Evaluation set & 15 RDKit path fits & Diverse \textsc{STORM} \\
\midrule
CASP16 & $0.6613$ & $0.6674 \pm 0.0086$ \\
FEP+4 & $0.6569$ & $0.6944 \pm 0.0230$ \\
OpenFE & $0.5055$ & $0.5814 \pm 0.0089$ \\
ChEMBL35 Time Split & $0.2562$ & $0.3529 \pm 0.0040$ \\
\bottomrule
\end{tabular}
\end{table}

\subsection{Matched protein-input and ligand-diversity controls}
\label{app:matched-ablation}

We compared four conditions using the same training corpus, objective, assay
endpoint inputs, shared checkpoint epoch 45, and graph-smoothing rule. Each
condition had 15 independently trained MLPs and 18 votes per ensemble seed.
The matched ligand-only control retained the ligand input and vote weight
of each submitted slot. For each ensemble seed, it reused the four
target-blind components and replaced the eleven target-aware slots with
independently trained ligand-and-endpoint models. The repeated target-aware
control uses 15 fits of a single target-aware component. The repeated
ligand-only control uses 15 fits of the target-blind RDKit path-count model,
the highest-scoring standalone component by mean Pearson correlation on
OpenFE and ChEMBL35. The original equal-average control in
Table~\ref{tab:storm-diversity-ablation} is a separate experiment.

We scored ensemble seeds 2023, 2024, and 2025 independently and averaged
their compound-weighted within-assay Pearson correlations. For the
uncertainty estimates, we used 50,000 within-assay bootstrap replicates,
sharing each draw across all conditions and seeds. Within each replicate
we averaged the three seed-level scores. Their standard deviation across
replicates gives the SE of the three-fit mean. We used paired bootstrap-$t$
intervals and applied Holm correction across 16 contrasts. The analysis
conditions on the fitted models, observed assays, and selected recipe.

The STORM-minus-matched contrast estimates the contribution of protein
inputs for this recipe. The matched-minus-repeated ligand-only contrast
estimates the contribution of ligand representation diversity within
ligand-only models. Both gains survived adjustment on ChEMBL35. On
OpenFE, the protein-input gain survived adjustment and the ligand-diversity
contrast was inconclusive. Both contrasts were inconclusive on FEP+4
and CASP16.

\begin{table}[p]
\centering
\small
\setlength{\tabcolsep}{3pt}
\caption{Compound-weighted within-assay Pearson $r$ across three independently fitted seeds. Values are means $\pm$ one bootstrap SE.}
\label{tab:matched-ablation-scores}
\begin{tabular}{@{}lcccc@{}}
\toprule
Condition & ChEMBL35 & OpenFE & FEP+4 & CASP16 \\
\midrule
STORM & $0.3530 \pm 0.0069$ & $0.5813 \pm 0.0229$ & $0.6941 \pm 0.0538$ & $0.6678 \pm 0.0446$ \\
Matched ligand-only STORM & $0.2910 \pm 0.0070$ & $0.5349 \pm 0.0240$ & $0.6727 \pm 0.0551$ & $0.6515 \pm 0.0455$ \\
Repeated target-aware MLP & $0.2917 \pm 0.0074$ & $0.5123 \pm 0.0259$ & $0.6525 \pm 0.0598$ & $0.6222 \pm 0.0540$ \\
Repeated ligand-only MLP & $0.2499 \pm 0.0072$ & $0.5011 \pm 0.0238$ & $0.6445 \pm 0.0543$ & $0.6537 \pm 0.0458$ \\
\bottomrule
\end{tabular}
\par\medskip
\setlength{\tabcolsep}{3pt}
\caption{Paired three-fit ablation contrasts. Intervals are unadjusted 95\% bootstrap-$t$ intervals. Holm adjustment covers all 16 tests.}
\label{tab:matched-ablation-tests}
\begin{tabular}{@{}llcr@{}}
\toprule
Benchmark & Contrast & $\Delta r$ [95\% CI] & Holm $p$ \\
\midrule
ChEMBL35 & STORM vs. matched ligand-only & $+0.0619\;[+0.0554,\,+0.0707]$ & $6.4\times10^{-4}$ \\
 & STORM vs. repeated target-aware & $+0.0612\;[+0.0522,\,+0.0728]$ & $6.4\times10^{-4}$ \\
 & STORM vs. repeated ligand-only & $+0.1031\;[+0.0937,\,+0.1157]$ & $6.4\times10^{-4}$ \\
 & Matched vs. repeated ligand-only & $+0.0411\;[+0.0327,\,+0.0508]$ & $6.4\times10^{-4}$ \\
\addlinespace
OpenFE & STORM vs. matched ligand-only & $+0.0464\;[+0.0191,\,+0.0658]$ & $0.00616$ \\
 & STORM vs. repeated target-aware & $+0.0690\;[+0.0262,\,+0.1112]$ & $0.0188$ \\
 & STORM vs. repeated ligand-only & $+0.0802\;[+0.0499,\,+0.1198]$ & $6.4\times10^{-4}$ \\
 & Matched vs. repeated ligand-only & $+0.0338\;[+0.0083,\,+0.0755]$ & $0.1267$ \\
\addlinespace
FEP+4 & STORM vs. matched ligand-only & $+0.0215\;[-0.0144,\,+0.0575]$ & $1$ \\
 & STORM vs. repeated target-aware & $+0.0416\;[-0.0593,\,+0.1285]$ & $1$ \\
 & STORM vs. repeated ligand-only & $+0.0497\;[-0.0007,\,+0.1111]$ & $0.4307$ \\
 & Matched vs. repeated ligand-only & $+0.0282\;[-0.0169,\,+0.0781]$ & $1$ \\
\addlinespace
CASP16 & STORM vs. matched ligand-only & $+0.0163\;[-0.0105,\,+0.0449]$ & $1$ \\
 & STORM vs. repeated target-aware & $+0.0456\;[-0.0390,\,+0.1157]$ & $1$ \\
 & STORM vs. repeated ligand-only & $+0.0141\;[-0.0352,\,+0.0653]$ & $1$ \\
 & Matched vs. repeated ligand-only & $-0.0022\;[-0.0464,\,+0.0436]$ & $1$ \\
\bottomrule
\end{tabular}
\end{table}

\subsection{Effect of graph smoothing}
\label{app:smoothing-ablation}

We scored the same three 15-member, 18-vote fits on the same measurements
with and without graph smoothing. We recovered the unsmoothed vectors by
inverting the three-pass linear smoothing operator within each assay using
the original Klekota--Roth fingerprint cache and activity-file row order,
preserving the original order for tied similarities. The recovered
seed-2023 vector agreed with the saved unsmoothed predictions within
$8\times10^{-15}$ across 16,404 rows.

We resampled rows within assays 50,000 times with NumPy seed 20261112,
sharing draws across both conditions and all three fits. We redrew
undefined Pearson correlations jointly and averaged the three seed-level
scores in each replicate. The paired percentile intervals condition on
the original graphs, fitted models, and observed assays. The intervals
exclude zero on ChEMBL35 and OpenFE.

\begin{table}[!htbp]
\centering
\small
\caption{Graph smoothing ablation. Scores are three-fit means $\pm$ one bootstrap SE. Differences are smoothed minus unsmoothed, with unadjusted 95\% paired percentile intervals.}
\label{tab:smoothing-ablation}
\begin{tabular}{@{}lccc@{}}
\toprule
Benchmark & Unsmoothed & Smoothed & $\Delta r$ [95\% CI] \\
\midrule
ChEMBL35 & $0.3491 \pm 0.0070$ & $0.3530 \pm 0.0069$ & $+0.0038\;[+0.0020,\,+0.0059]$ \\
OpenFE & $0.5708 \pm 0.0233$ & $0.5813 \pm 0.0228$ & $+0.0105\;[+0.0031,\,+0.0182]$ \\
FEP+4 & $0.6832 \pm 0.0568$ & $0.6941 \pm 0.0538$ & $+0.0109\;[-0.0195,\,+0.0387]$ \\
CASP16 & $0.6616 \pm 0.0450$ & $0.6678 \pm 0.0446$ & $+0.0062\;[-0.0109,\,+0.0247]$ \\
\bottomrule
\end{tabular}
\end{table}

\section{Per-class ChEMBL35 Time Split Results}
Table~\ref{tab:chembl35_classes} gives compound-weighted within-assay
Pearson correlations for the six ChEMBL35 target classes. The benchmark
has 449 assays across 278 distinct protein IDs. The parenthetical counts
in the table refer to assays.

The published Boltz-2 references on OpenFE, FEP+4, and CASP16 come from
Figure~6 of \citet{passaro2025boltz2}, whose evaluation
weights assays by compound count and lists 876 OpenFE measurements.
The approximately 0.341 overall ChEMBL35 reference is our reconstruction
from the per-class Boltz-2 results in IsoDDE Figure~8~\citep{isodde2026}.
We weighted each reported class score by its compound count. IsoDDE
leaves within-class weighting unspecified. We estimated published error
bars from the source figures for descriptive comparison. Statistical
claims about Boltz-2 use the local matched analysis in
Appendix~\ref{app:matched-benchmark-evaluation}.

\begin{table}[!htbp]
\centering
\resizebox{\textwidth}{!}{%
\begin{tabular}{lccccccc}
\hline
Model
& Overall ($n=449$)
& Enzyme ($n=135$)
& GPCR ($n=97$)
& Ion Channel ($n=17$)
& Kinase ($n=115$)
& Other ($n=51$)
& Transcription Factor ($n=34$) \\
\hline
\textbf{STORM}
& $\mathbf{0.353 \pm 0.004}$
& $\mathbf{0.318 \pm 0.012}$
& $\mathbf{0.412 \pm 0.012}$
& $\underline{0.319 \pm 0.019}$
& $\underline{0.352 \pm 0.009}$
& $\mathbf{0.337 \pm 0.013}$
& $\mathbf{0.348 \pm 0.015}$ \\
Boltz-2
& $\underline{0.341 \pm 0.009}$
& $\underline{0.307 \pm 0.017}$
& $\underline{0.298 \pm 0.019}$
& $0.143 \pm 0.047$
& $\mathbf{0.416 \pm 0.018}$
& $\underline{0.336 \pm 0.028}$
& $\underline{0.347 \pm 0.033}$ \\
Structure-free DTA
& $0.235 \pm 0.013$
& $0.178 \pm 0.027$
& $0.292 \pm 0.030$
& $\mathbf{0.331 \pm 0.061}$
& $0.217 \pm 0.019$
& $0.258 \pm 0.046$
& $0.284 \pm 0.068$ \\
AttentionDTA
& $0.154 \pm 0.014$
& $0.108 \pm 0.030$
& $0.263 \pm 0.027$
& $0.098 \pm 0.057$
& $0.110 \pm 0.020$
& $0.187 \pm 0.043$
& $0.179 \pm 0.059$ \\
PSICHIC
& $0.139 \pm 0.015$
& $0.132 \pm 0.033$
& $0.155 \pm 0.038$
& $0.198 \pm 0.071$
& $0.151 \pm 0.020$
& $0.104 \pm 0.049$
& $0.074 \pm 0.071$ \\
BALM
& $0.209 \pm 0.014$
& $0.198 \pm 0.029$
& $0.254 \pm 0.027$
& $0.225 \pm 0.066$
& $0.186 \pm 0.023$
& $0.238 \pm 0.043$
& $0.178 \pm 0.062$ \\
Mattsson-Walters
& $0.228 \pm 0.014$
& $0.227 \pm 0.029$
& $0.297 \pm 0.033$
& $0.113 \pm 0.087$
& $0.202 \pm 0.022$
& $0.207 \pm 0.049$
& $0.245 \pm 0.056$ \\
\hline
\end{tabular}%
}
\caption{Compound-weighted within-assay Pearson correlation on the
ChEMBL35 Time Split benchmark. Parenthetical $n$ values count assays. Boltz-2 values are reported references.}
\label{tab:chembl35_classes}
\end{table}

\section{Estimating L4-equivalent GPU hours needed to Pretrain Boltz-2}
\label{app:boltz-compute}

Figure~\ref{fig:speed} uses the no-MSA Boltz-2 inference timing measured on
an NVIDIA L4 (Section~\ref{sec:fast inference}). The benchmark accuracy
evaluation used ColabFold MSAs
(Appendix~\ref{app:matched-benchmark-evaluation}). The training panel reports
an estimate of the L4-equivalent GPU hours needed to train Boltz-2 from scratch.

We use OpenFold3-preview 2 \citep{openfold3-preview}, a comparable cofolding only method, as our reference point, as some training details are available. The model was trained for 14 days on 256 NVIDIA H100 GPUs, leading to 86,016 H100 GPU hours. OpenFold3-preview 2 was trained on 13M structures, and Boltz-2 was trained on 5M structures (both include distilled structures). For simplicity, we assume a linear relationship between compute and the size of the distilled training dataset (this may not always be the case). Finally, H100s have a peak BF16 Tensor Core throughput of 1,979 TFLOP/s \citep{nvidia_h100_specs}, compared to the L4's 242 TFLOP/s \citep{nvidia_l4_specs}. Multiplying the 86,016 H100 GPU hours by the TFLOP/s and training dataset sizes gives us

$$86,016 \times \frac{1,979}{242} \times \frac{5}{13} \approx 270,000 \text{ L4-equivalent GPU hours}$$

We once again emphasize that this is a very rough estimation, and many details about Boltz-2's runtime are not available publicly. We intend this estimate only to indicate the approximate order of magnitude of Boltz-2's training compute.

\section{Experimental Setup}

\subsection{Additional Details on Finetuning Study}
\label{app:finetuning-setup}

We evaluated finetuning on three targets absent from the base model's training set. For LpxH, we used the larger of two nonoverlapping assays as the support set (92 compounds) and the smaller assay as the held-out test set (42 compounds). A0A0B5JFF1 had one eligible assay with 80 compounds; a fixed random compound split assigned 56 to support and 24 to test. For ALPK1, we used a within-assay demonstration split rather than the available assay pair. We clustered compounds using 2,048-bit radius-2 Morgan fingerprints and RDKit Butina clustering with a Tanimoto distance cutoff of 0.5, then selected 80 support and 40 test compounds across the clusters. The remaining 318 eligible compounds were not used. This procedure sampled compounds using the clusters; it did not hold out entire clusters.

For each target, we selected finetuning settings using three independent 80/20 partitions of the support compounds. We trained on each partition's training portion and ranked settings by mean Pearson correlation on its held-out support portion, using the canonical Morgan component from pretrained seed 2023 for the search. The test compounds were excluded from this selection. We then restarted all 15 components from their epoch-45 pretrained checkpoints, finetuned each component on the complete support set for each of the three pretrained seeds, and combined their predictions using the original 18 votes. Full-model finetuning used a learning rate of $10^{-5}$ and an L2-SP penalty weight of $0.1$ for all three targets, with 50 steps for LpxH, 100 for A0A0B5JFF1, and 200 for ALPK1. We compared each adapted ensemble with its corresponding untouched base ensemble on the same held-out test compounds.

\subsection{Finetuning Loss}
For each target, we initialized the model from its pretrained checkpoint and
fine-tuned it on the support assay. We updated all model parameters, minimizing

\begin{equation}
\mathcal{L}_{\mathrm{FT}}(\theta)
=
\frac{1}{N}\sum_{i=1}^{N}
\operatorname{SmoothL1}\!\left(f_{\theta}(x_i),y_i\right)
+
\lambda \sum_{p\in\mathcal{P}}
\left\|\theta_p-\theta^{(0)}_p\right\|_2^2,
\end{equation}
where $\theta^{(0)}$ denotes the pretrained parameters and $\mathcal{P}$
contains the parameters updated in the chosen adaptation mode. The second term
is L2-SP regularization, which penalizes movement away from the pretrained
starting point \citep{pmlr-v80-li18a}. The penalty is summed over parameters
without normalization.

We selected the number of update steps and $\lambda$ using validation
partitions of the support assay only. We then restarted from the pretrained
checkpoint, fine-tuned on all support compounds using the selected settings,
and evaluated once on the held-out test assay. Test-assay labels were not used
for adaptation or hyperparameter selection.

\subsection{Laptop Training Benchmark}
\label{app:laptop-training}

We benchmarked training the 15-component ensemble with seed 2023 on a
2024 MacBook Pro with an Apple M4 Pro and 24~GB of unified memory. We ran
four concurrent CPU fits with two threads per fit, or two concurrent fits
using the PyTorch MPS backend. For each configuration, we timed two runs
of three epochs per component using prepared features. The wall-clock
measurements covered the training job queue from launch to completion after feature preparation.

For each component, we added 42 times the mean duration of epochs 2 and 3
(including epoch preparation) to the measured job duration, then replayed
the scheduling policy to estimate the 45-epoch training time. Mean projected
wall times across the two runs were 2~h 04~min 36~s on CPU (SE 4~min 29~s)
and 1~h 20~min 23~s on MPS (SE 8~min 21~s).

\clearpage
\section{Matched Benchmark Evaluation}
\label{app:matched-benchmark-evaluation}

\subsection{Cohort and Boltz-2 inference}

We compared the submitted, voted and graph-smoothed seed-2023 \textsc{STORM}
ensemble, Boltz-2 inference with seed 2023, and the five retrained
structure-free baselines (Structure-free DTA, AttentionDTA, PSICHIC, BALM,
and Mattsson--Walters) on the same compound measurements. We joined the
prediction exports by benchmark, assay (series ID), and activity ID and checked
the standardized ligands and experimental labels across the exports. The
\textsc{STORM} graphs were built on the original assays before restricting the
scored rows.

For the accuracy evaluation, we supplied ColabFold MSAs to Boltz-2
v2.2.1 and used the default inference settings in the
\href{https://github.com/jwohlwend/boltz/blob/v2.2.1/docs/prediction.md\#options}{Boltz-2 prediction guide}.
For scoring, we applied the release's optional molecular-weight correction
(\texttt{--affinity\_mw\_correction}) to the saved affinity values using
the released coefficients. The correction is off by default. We applied no
calibration fitted to the benchmark labels. Only the inference speed measurement in Section~\ref{sec:fast inference} used
Boltz-2 without MSAs.

Boltz-2 failed to produce predictions for 159 ChEMBL35 measurements, covering
106 distinct protein--ligand pairs across 23 assays. We excluded these
measurements from scoring for every model. Four ChEMBL35 assays had no
successful Boltz-2 predictions and were omitted entirely. The matched
comparison therefore includes 15,145 measurements from 445 of the original
449 assays. The 426 assays with complete coverage contain 14,594 measurements.
The sequence-identity analysis uses a minimum of three matched measurements
per assay, which leaves 443 ChEMBL35 assays.

\begin{table}[htbp]
\centering
\small
\caption{Measurement and assay coverage in the matched benchmark comparison.}
\label{tab:matched-benchmark-coverage}
\begin{tabular}{@{}lrrrr@{}}
\toprule
Benchmark & All rows & Matched rows & All assays & Matched assays \\
\midrule
ChEMBL35 Time Split & 15,304 & 15,145 & 449 & 445 \\
OpenFE & 873 & 873 & 58 & 58 \\
FEP+4 & 87 & 87 & 4 & 4 \\
CASP16 & 140 & 140 & 2 & 2 \\
\bottomrule
\end{tabular}
\end{table}

\subsection{Metrics and paired inference}

We computed Pearson correlation separately within each assay and averaged
the correlations using the number of matched measurements as weights
\citep{passaro2025boltz2}. We also computed pooled MAE and RMSE in pActivity
units. Each matched measurement received equal weight. For RMSE, we took the
square root after averaging squared errors across all matched measurements.
ChEMBL35 uses the recorded pActivity labels. For OpenFE, FEP+4, and CASP16,
we converted experimental free energies as $-\Delta G/(RT\ln 10)$ at
298.15~K and oriented the structure-free exports so higher values indicate
stronger binding. We converted the Boltz-2 log-micromolar affinity output
$x$ to $6-x$. Boltz-2 pIC50 serves as a proxy for the mixed ChEMBL35 endpoints.

For each benchmark we drew 50,000 bootstrap replicates by sampling the
original number of matched rows with replacement within each assay. The
same row draws were used for all seven models. If a draw produced an undefined
Pearson correlation for any model, we redrew that assay jointly for Pearson.
The score SE is the sample standard deviation across replicates. For each
STORM-minus-comparator contrast, we used an assay-stratified SE based on
paired absolute errors for MAE, the paired squared-error covariance and the
square-root delta method for RMSE, and the paired correlation influence
values for Pearson. We recomputed the SE in each replicate and used the
studentized distribution for equal-tailed 95\%
bootstrap-$t$ intervals and two-sided $p$-values. We doubled the smaller
tail probability with a plus-one Monte Carlo correction and applied Holm
correction jointly across four benchmarks, three metrics, and six comparators.
The analysis conditions on the observed assays, fitted predictions, and
original smoothing graphs. The two inconclusive Pearson contrasts against
structure-free baselines were Mattsson--Walters on FEP+4 and AttentionDTA
on CASP16. All statistical claims about Boltz-2 use this
local paired evaluation. The published Boltz-2 uncertainty bars were
estimated from the figures in the Boltz-2 and IsoDDE reports and are shown
for descriptive comparison \citep{passaro2025boltz2,isodde2026}.

Tables~\ref{tab:matched-pearson}--\ref{tab:matched-rmse} give the matched
model scores. Tables~\ref{tab:paired-isodde_chembl35}--\ref{tab:paired-casp16}
report all 72 contrasts. On ChEMBL35, the paired Pearson difference between
\textsc{STORM} and corrected Boltz-2 was $+0.0521$ (95\% CI
$[+0.0359,+0.0691]$, Holm-adjusted $p=0.00288$). On the 426 complete
ChEMBL35 assays, the Pearson point estimates were 0.3583 for \textsc{STORM}
and 0.3008 for corrected Boltz-2. The Pearson differences against corrected
Boltz-2 were inconclusive on OpenFE, FEP+4, and CASP16 after adjustment.

\begin{table}[p]
\centering
\small
\caption{Matched Pearson $r$. Values are estimates $\pm$ one within-assay bootstrap SE. Bold marks the best point estimate.}
\label{tab:matched-pearson}
\setlength{\tabcolsep}{2pt}
\begin{tabular}{@{}lcccc@{}}
\toprule
Model & ChEMBL35 & OpenFE & FEP+4 & CASP16 \\
\midrule
STORM & $\mathbf{0.3549 \pm 0.0072}$ & $\mathbf{0.5762 \pm 0.0239}$ & $\mathbf{0.7048 \pm 0.0533}$ & $\mathbf{0.6706 \pm 0.0443}$ \\
Boltz-2 (MW corrected) & $0.3028 \pm 0.0074$ & $0.5696 \pm 0.0236$ & $0.6753 \pm 0.0616$ & $0.6403 \pm 0.0521$ \\
Structure-free DTA & $0.2378 \pm 0.0076$ & $0.3283 \pm 0.0304$ & $0.4413 \pm 0.0923$ & $0.3801 \pm 0.0753$ \\
AttentionDTA & $0.1572 \pm 0.0078$ & $0.2439 \pm 0.0314$ & $0.0617 \pm 0.0893$ & $0.5803 \pm 0.0486$ \\
PSICHIC & $0.1424 \pm 0.0076$ & $0.3810 \pm 0.0286$ & $0.5178 \pm 0.0724$ & $0.5651 \pm 0.0537$ \\
BALM & $0.2095 \pm 0.0077$ & $0.3284 \pm 0.0308$ & $0.4807 \pm 0.0860$ & $0.3867 \pm 0.0662$ \\
Mattsson--Walters & $0.2295 \pm 0.0077$ & $0.2637 \pm 0.0319$ & $0.6382 \pm 0.0728$ & $0.2518 \pm 0.0852$ \\
\bottomrule
\end{tabular}
\par\medskip
\caption{Matched MAE in pActivity units. Values are estimates $\pm$ one within-assay bootstrap SE. Bold marks the best point estimate.}
\label{tab:matched-mae}
\setlength{\tabcolsep}{2pt}
\begin{tabular}{@{}lcccc@{}}
\toprule
Model & ChEMBL35 & OpenFE & FEP+4 & CASP16 \\
\midrule
STORM & $\mathbf{0.8600 \pm 0.0042}$ & $1.0436 \pm 0.0186$ & $0.9476 \pm 0.0558$ & $0.8741 \pm 0.0386$ \\
Boltz-2 (MW corrected) & $1.0779 \pm 0.0052$ & $\mathbf{0.8559 \pm 0.0172}$ & $\mathbf{0.5541 \pm 0.0425}$ & $\mathbf{0.7814 \pm 0.0505}$ \\
Structure-free DTA & $0.9665 \pm 0.0049$ & $1.3962 \pm 0.0233$ & $1.2331 \pm 0.0694$ & $1.0288 \pm 0.0506$ \\
AttentionDTA & $0.9037 \pm 0.0047$ & $1.1244 \pm 0.0213$ & $0.8718 \pm 0.0661$ & $0.8306 \pm 0.0415$ \\
PSICHIC & $0.9719 \pm 0.0046$ & $1.1207 \pm 0.0203$ & $0.7050 \pm 0.0507$ & $0.9148 \pm 0.0437$ \\
BALM & $0.9059 \pm 0.0046$ & $1.2255 \pm 0.0210$ & $0.9190 \pm 0.0591$ & $1.0384 \pm 0.0493$ \\
Mattsson--Walters & $0.9421 \pm 0.0048$ & $1.2406 \pm 0.0214$ & $1.2269 \pm 0.0552$ & $1.6323 \pm 0.0828$ \\
\bottomrule
\end{tabular}
\par\medskip
\caption{Matched RMSE in pActivity units. Values are estimates $\pm$ one within-assay bootstrap SE. Bold marks the best point estimate.}
\label{tab:matched-rmse}
\setlength{\tabcolsep}{2pt}
\begin{tabular}{@{}lcccc@{}}
\toprule
Model & ChEMBL35 & OpenFE & FEP+4 & CASP16 \\
\midrule
STORM & $\mathbf{1.0663 \pm 0.0047}$ & $1.2920 \pm 0.0235$ & $1.1743 \pm 0.0609$ & $1.0110 \pm 0.0393$ \\
Boltz-2 (MW corrected) & $1.3554 \pm 0.0061$ & $\mathbf{1.1085 \pm 0.0244}$ & $\mathbf{0.7022 \pm 0.0519}$ & $0.9865 \pm 0.0594$ \\
Structure-free DTA & $1.2321 \pm 0.0057$ & $1.7102 \pm 0.0258$ & $1.4457 \pm 0.0736$ & $1.2071 \pm 0.0538$ \\
AttentionDTA & $1.1515 \pm 0.0054$ & $1.4013 \pm 0.0256$ & $1.0761 \pm 0.0730$ & $\mathbf{0.9860 \pm 0.0437}$ \\
PSICHIC & $1.1961 \pm 0.0050$ & $1.3721 \pm 0.0251$ & $0.8590 \pm 0.0532$ & $1.0608 \pm 0.0440$ \\
BALM & $1.1485 \pm 0.0054$ & $1.4933 \pm 0.0257$ & $1.0957 \pm 0.0641$ & $1.2250 \pm 0.0486$ \\
Mattsson--Walters & $1.2189 \pm 0.0059$ & $1.5260 \pm 0.0282$ & $1.3938 \pm 0.0549$ & $1.9066 \pm 0.0766$ \\
\bottomrule
\end{tabular}
\end{table}

\begin{table}[p]
\centering
\small
\caption{Paired STORM-minus-comparator tests on ChEMBL35 Time Split. Intervals are unadjusted 95\% bootstrap-$t$ intervals. Bold adjusted $p$-values are below 0.05 after Holm correction across all 72 contrasts.}
\label{tab:paired-isodde_chembl35}
\begin{tabular}{@{}llcrr@{}}
\toprule
Metric & Comparator & $\Delta$ [95\% CI] & Raw $p$ & Holm $p$ \\
\midrule
Pearson $r$ & Boltz-2 (MW corrected) & $+0.0521\;[+0.0359,\,+0.0691]$ & $4.00\times10^{-5}$ & $\mathbf{0.00288}$ \\
 & Structure-free DTA & $+0.1171\;[+0.1011,\,+0.1331]$ & $4.00\times10^{-5}$ & $\mathbf{0.00288}$ \\
 & AttentionDTA & $+0.1977\;[+0.1815,\,+0.2173]$ & $4.00\times10^{-5}$ & $\mathbf{0.00288}$ \\
 & PSICHIC & $+0.2126\;[+0.1977,\,+0.2329]$ & $4.00\times10^{-5}$ & $\mathbf{0.00288}$ \\
 & BALM & $+0.1454\;[+0.1303,\,+0.1629]$ & $4.00\times10^{-5}$ & $\mathbf{0.00288}$ \\
 & Mattsson--Walters & $+0.1254\;[+0.1093,\,+0.1415]$ & $4.00\times10^{-5}$ & $\mathbf{0.00288}$ \\
MAE & Boltz-2 (MW corrected) & $-0.2179\;[-0.2288,\,-0.2070]$ & $4.00\times10^{-5}$ & $\mathbf{0.00288}$ \\
 & Structure-free DTA & $-0.1065\;[-0.1150,\,-0.0981]$ & $4.00\times10^{-5}$ & $\mathbf{0.00288}$ \\
 & AttentionDTA & $-0.0437\;[-0.0509,\,-0.0364]$ & $4.00\times10^{-5}$ & $\mathbf{0.00288}$ \\
 & PSICHIC & $-0.1119\;[-0.1172,\,-0.1065]$ & $4.00\times10^{-5}$ & $\mathbf{0.00288}$ \\
 & BALM & $-0.0459\;[-0.0523,\,-0.0394]$ & $4.00\times10^{-5}$ & $\mathbf{0.00288}$ \\
 & Mattsson--Walters & $-0.0821\;[-0.0900,\,-0.0742]$ & $4.00\times10^{-5}$ & $\mathbf{0.00288}$ \\
RMSE & Boltz-2 (MW corrected) & $-0.2891\;[-0.3010,\,-0.2772]$ & $4.00\times10^{-5}$ & $\mathbf{0.00288}$ \\
 & Structure-free DTA & $-0.1658\;[-0.1747,\,-0.1571]$ & $4.00\times10^{-5}$ & $\mathbf{0.00288}$ \\
 & AttentionDTA & $-0.0853\;[-0.0931,\,-0.0775]$ & $4.00\times10^{-5}$ & $\mathbf{0.00288}$ \\
 & PSICHIC & $-0.1299\;[-0.1351,\,-0.1247]$ & $4.00\times10^{-5}$ & $\mathbf{0.00288}$ \\
 & BALM & $-0.0823\;[-0.0891,\,-0.0755]$ & $4.00\times10^{-5}$ & $\mathbf{0.00288}$ \\
 & Mattsson--Walters & $-0.1526\;[-0.1616,\,-0.1438]$ & $4.00\times10^{-5}$ & $\mathbf{0.00288}$ \\
\bottomrule
\end{tabular}
\par\medskip
\caption{Paired STORM-minus-comparator tests on OpenFE. Intervals are unadjusted 95\% bootstrap-$t$ intervals. Bold adjusted $p$-values are below 0.05 after Holm correction across all 72 contrasts.}
\label{tab:paired-openfe}
\begin{tabular}{@{}llcrr@{}}
\toprule
Metric & Comparator & $\Delta$ [95\% CI] & Raw $p$ & Holm $p$ \\
\midrule
Pearson $r$ & Boltz-2 (MW corrected) & $+0.0066\;[-0.0516,\,+0.0560]$ & $0.9285$ & $1$ \\
 & Structure-free DTA & $+0.2479\;[+0.1774,\,+0.3172]$ & $4.00\times10^{-5}$ & $\mathbf{0.00288}$ \\
 & AttentionDTA & $+0.3323\;[+0.2685,\,+0.3976]$ & $4.00\times10^{-5}$ & $\mathbf{0.00288}$ \\
 & PSICHIC & $+0.1952\;[+0.1393,\,+0.2535]$ & $4.00\times10^{-5}$ & $\mathbf{0.00288}$ \\
 & BALM & $+0.2478\;[+0.1786,\,+0.3207]$ & $4.00\times10^{-5}$ & $\mathbf{0.00288}$ \\
 & Mattsson--Walters & $+0.3124\;[+0.2258,\,+0.3823]$ & $4.00\times10^{-5}$ & $\mathbf{0.00288}$ \\
MAE & Boltz-2 (MW corrected) & $+0.1877\;[+0.1403,\,+0.2347]$ & $4.00\times10^{-5}$ & $\mathbf{0.00288}$ \\
 & Structure-free DTA & $-0.3526\;[-0.3859,\,-0.3180]$ & $4.00\times10^{-5}$ & $\mathbf{0.00288}$ \\
 & AttentionDTA & $-0.0808\;[-0.1095,\,-0.0526]$ & $4.00\times10^{-5}$ & $\mathbf{0.00288}$ \\
 & PSICHIC & $-0.0771\;[-0.0948,\,-0.0589]$ & $4.00\times10^{-5}$ & $\mathbf{0.00288}$ \\
 & BALM & $-0.1819\;[-0.2067,\,-0.1566]$ & $4.00\times10^{-5}$ & $\mathbf{0.00288}$ \\
 & Mattsson--Walters & $-0.1970\;[-0.2274,\,-0.1663]$ & $4.00\times10^{-5}$ & $\mathbf{0.00288}$ \\
RMSE & Boltz-2 (MW corrected) & $+0.1835\;[+0.1327,\,+0.2348]$ & $4.00\times10^{-5}$ & $\mathbf{0.00288}$ \\
 & Structure-free DTA & $-0.4182\;[-0.4519,\,-0.3834]$ & $4.00\times10^{-5}$ & $\mathbf{0.00288}$ \\
 & AttentionDTA & $-0.1093\;[-0.1371,\,-0.0820]$ & $4.00\times10^{-5}$ & $\mathbf{0.00288}$ \\
 & PSICHIC & $-0.0801\;[-0.0986,\,-0.0621]$ & $4.00\times10^{-5}$ & $\mathbf{0.00288}$ \\
 & BALM & $-0.2013\;[-0.2246,\,-0.1774]$ & $4.00\times10^{-5}$ & $\mathbf{0.00288}$ \\
 & Mattsson--Walters & $-0.2340\;[-0.2647,\,-0.2036]$ & $4.00\times10^{-5}$ & $\mathbf{0.00288}$ \\
\bottomrule
\end{tabular}
\end{table}

\begin{table}[p]
\centering
\small
\caption{Paired STORM-minus-comparator tests on FEP+4. Intervals are unadjusted 95\% bootstrap-$t$ intervals. Bold adjusted $p$-values are below 0.05 after Holm correction across all 72 contrasts.}
\label{tab:paired-fep_plus_4p}
\begin{tabular}{@{}llcrr@{}}
\toprule
Metric & Comparator & $\Delta$ [95\% CI] & Raw $p$ & Holm $p$ \\
\midrule
Pearson $r$ & Boltz-2 (MW corrected) & $+0.0295\;[-0.0655,\,+0.1386]$ & $0.5327$ & $1$ \\
 & Structure-free DTA & $+0.2635\;[+0.1129,\,+0.4617]$ & $0.00128$ & $\mathbf{0.0256}$ \\
 & AttentionDTA & $+0.6431\;[+0.4932,\,+0.8051]$ & $4.00\times10^{-5}$ & $\mathbf{0.00288}$ \\
 & PSICHIC & $+0.1870\;[+0.0773,\,+0.2906]$ & $0.00172$ & $\mathbf{0.03096}$ \\
 & BALM & $+0.2241\;[+0.0832,\,+0.4238]$ & $0.00208$ & $\mathbf{0.03536}$ \\
 & Mattsson--Walters & $+0.0666\;[-0.0513,\,+0.1939]$ & $0.264$ & $1$ \\
MAE & Boltz-2 (MW corrected) & $+0.3936\;[+0.2317,\,+0.5405]$ & $8.00\times10^{-5}$ & $\mathbf{0.00288}$ \\
 & Structure-free DTA & $-0.2854\;[-0.4043,\,-0.1562]$ & $0.00012$ & $\mathbf{0.003}$ \\
 & AttentionDTA & $+0.0759\;[-0.0220,\,+0.1743]$ & $0.1245$ & $1$ \\
 & PSICHIC & $+0.2426\;[+0.1761,\,+0.2862]$ & $4.00\times10^{-5}$ & $\mathbf{0.00288}$ \\
 & BALM & $+0.0286\;[-0.0280,\,+0.0919]$ & $0.3238$ & $1$ \\
 & Mattsson--Walters & $-0.2793\;[-0.3688,\,-0.1565]$ & $8.00\times10^{-5}$ & $\mathbf{0.00288}$ \\
RMSE & Boltz-2 (MW corrected) & $+0.4722\;[+0.2851,\,+0.6402]$ & $4.00\times10^{-5}$ & $\mathbf{0.00288}$ \\
 & Structure-free DTA & $-0.2713\;[-0.3866,\,-0.1468]$ & $0.00012$ & $\mathbf{0.003}$ \\
 & AttentionDTA & $+0.0982\;[-0.0043,\,+0.2027]$ & $0.0604$ & $0.6644$ \\
 & PSICHIC & $+0.3154\;[+0.2577,\,+0.3620]$ & $4.00\times10^{-5}$ & $\mathbf{0.00288}$ \\
 & BALM & $+0.0787\;[+0.0164,\,+0.1340]$ & $0.01688$ & $0.2026$ \\
 & Mattsson--Walters & $-0.2195\;[-0.2994,\,-0.1271]$ & $4.00\times10^{-5}$ & $\mathbf{0.00288}$ \\
\bottomrule
\end{tabular}
\par\medskip
\caption{Paired STORM-minus-comparator tests on CASP16. Intervals are unadjusted 95\% bootstrap-$t$ intervals. Bold adjusted $p$-values are below 0.05 after Holm correction across all 72 contrasts.}
\label{tab:paired-casp16}
\begin{tabular}{@{}llcrr@{}}
\toprule
Metric & Comparator & $\Delta$ [95\% CI] & Raw $p$ & Holm $p$ \\
\midrule
Pearson $r$ & Boltz-2 (MW corrected) & $+0.0302\;[-0.0528,\,+0.1101]$ & $0.491$ & $1$ \\
 & Structure-free DTA & $+0.2905\;[+0.1436,\,+0.4306]$ & $0.00012$ & $\mathbf{0.003}$ \\
 & AttentionDTA & $+0.0903\;[+0.0219,\,+0.1645]$ & $0.00764$ & $0.107$ \\
 & PSICHIC & $+0.1055\;[+0.0412,\,+0.1817]$ & $0.00156$ & $\mathbf{0.02964}$ \\
 & BALM & $+0.2838\;[+0.1605,\,+0.4091]$ & $8.00\times10^{-5}$ & $\mathbf{0.00288}$ \\
 & Mattsson--Walters & $+0.4187\;[+0.2532,\,+0.5970]$ & $4.00\times10^{-5}$ & $\mathbf{0.00288}$ \\
MAE & Boltz-2 (MW corrected) & $+0.0927\;[-0.0070,\,+0.1904]$ & $0.06824$ & $0.6824$ \\
 & Structure-free DTA & $-0.1546\;[-0.2618,\,-0.0517]$ & $0.00292$ & $\mathbf{0.04672}$ \\
 & AttentionDTA & $+0.0436\;[-0.0278,\,+0.1156]$ & $0.2298$ & $1$ \\
 & PSICHIC & $-0.0406\;[-0.0723,\,-0.0084]$ & $0.01444$ & $0.1877$ \\
 & BALM & $-0.1643\;[-0.2479,\,-0.0836]$ & $0.00016$ & $\mathbf{0.00352}$ \\
 & Mattsson--Walters & $-0.7582\;[-0.9466,\,-0.5563]$ & $4.00\times10^{-5}$ & $\mathbf{0.00288}$ \\
RMSE & Boltz-2 (MW corrected) & $+0.0245\;[-0.0900,\,+0.1353]$ & $0.6785$ & $1$ \\
 & Structure-free DTA & $-0.1960\;[-0.3030,\,-0.0920]$ & $0.0002$ & $\mathbf{0.0042}$ \\
 & AttentionDTA & $+0.0250\;[-0.0405,\,+0.0910]$ & $0.4556$ & $1$ \\
 & PSICHIC & $-0.0498\;[-0.0827,\,-0.0171]$ & $0.0032$ & $\mathbf{0.048}$ \\
 & BALM & $-0.2140\;[-0.2966,\,-0.1345]$ & $4.00\times10^{-5}$ & $\mathbf{0.00288}$ \\
 & Mattsson--Walters & $-0.8956\;[-1.0647,\,-0.7155]$ & $4.00\times10^{-5}$ & $\mathbf{0.00288}$ \\
\bottomrule
\end{tabular}
\end{table}

\subsection{Molecular-weight correction and resampling sensitivity}
\label{app:matched-sensitivity}

Table~\ref{tab:boltz-mw-sensitivity} compares the default and corrected
Boltz-2 outputs on identical measurements. The correction increased
within-assay Pearson correlation on all four benchmarks. It reduced both
absolute-error point estimates on OpenFE, FEP+4, and CASP16 and increased
both on ChEMBL35.

As a sensitivity analysis, we resampled complete assays for MAE and RMSE,
keeping all matched measurements in each selected assay. We applied Holm
correction across the eight absolute-error sensitivity tests. Only the two
ChEMBL35 differences remained significant. For Pearson we resampled
UniProt targets with replacement and kept their scored assays intact.
Table~\ref{tab:matched-resampling-sensitivity} reports the unadjusted
percentile intervals. FEP+4 and CASP16 contribute only four and two targets,
respectively.

\begin{table}[!htbp]
\centering
\small
\caption{Boltz-2 point estimates with default settings and with molecular-weight correction, shown as default/corrected on the same records.}
\label{tab:boltz-mw-sensitivity}
\begin{tabular}{@{}lccc@{}}
\toprule
Benchmark & Pearson $r$ & MAE & RMSE \\
\midrule
ChEMBL35 & $0.2909/0.3028$ & $0.9432/1.0779$ & $1.1993/1.3554$ \\
OpenFE & $0.5519/0.5696$ & $0.9601/0.8559$ & $1.2723/1.1085$ \\
FEP+4 & $0.6203/0.6753$ & $0.6462/0.5541$ & $0.7981/0.7022$ \\
CASP16 & $0.5920/0.6403$ & $0.8162/0.7814$ & $1.0188/0.9865$ \\
\bottomrule
\end{tabular}
\par\medskip
\setlength{\tabcolsep}{3pt}
\caption{Resampling sensitivity for STORM minus corrected Boltz-2. The Pearson column gives the target-bootstrap 95\% interval. MAE and RMSE give the difference and whole-assay bootstrap 95\% interval, with Holm $p$-values.}
\label{tab:matched-resampling-sensitivity}
\resizebox{\linewidth}{!}{%
\begin{tabular}{@{}lccrcr@{}}
\toprule
Benchmark & $\Delta r$ CI & $\Delta$ MAE [CI] & $p$ & $\Delta$ RMSE [CI] & $p$ \\
\midrule
ChEMBL35 & $[+0.0202,\,+0.0845]$ & $-0.2179\;[-0.2738,\,-0.1617]$ & $1.6\times10^{-4}$ & $-0.2891\;[-0.3537,\,-0.2247]$ & $1.6\times10^{-4}$ \\
OpenFE & $[-0.0695,\,+0.0810]$ & $+0.1877\;[-0.0033,\,+0.3851]$ & $0.3481$ & $+0.1835\;[-0.0191,\,+0.4011]$ & $0.3794$ \\
FEP+4 & $[-0.1104,\,+0.1254]$ & $+0.3936\;[-0.0797,\,+0.7715]$ & $0.4324$ & $+0.4722\;[-0.0788,\,+0.8039]$ & $0.3794$ \\
CASP16 & $[-0.1406,\,+0.0539]$ & $+0.0927\;[-0.0189,\,+0.9006]$ & $0.9986$ & $+0.0245\;[-0.1159,\,+0.9143]$ & $0.9986$ \\
\bottomrule
\end{tabular}%
}
\end{table}

\begin{table}[htbp]
\centering
\small
\caption{AttentionDTA Pearson correlations on FEP+4. The overall SE uses
compound resampling within the four fixed assays.}
\label{tab:attentiondta-fep}
\begin{tabular}{@{}lrr@{}}
\toprule
Target & Compounds & Pearson $r$ \\
\midrule
CDK2 & 16 & $-0.785$ \\
JNK1 & 21 & $0.003$ \\
p38$\alpha$ & 34 & $0.259$ \\
TYK2 & 16 & $0.567$ \\
\midrule
Overall & 87 & $0.062 \pm 0.089$ \\
\bottomrule
\end{tabular}
\end{table}

The negative CDK2 correlation contributes to the low overall score.
Omitting CDK2 gives a compound-weighted correlation of 0.253 across the
remaining three assays. Resampling whole assays gives an SE of 0.215,
close to the 0.218 reported in the original submission.

\clearpage
\section{Similarity to the Training Corpus}
\label{app:similarity}

We stratified the matched seed-2023 STORM and corrected Boltz-2 predictions
by proximity to the retained STORM affinity training corpus. This shared
reference applies to both models. Boltz-2 used additional affinity sources
whose exact overlap with the benchmarks is unknown. STORM graph smoothing
was applied to the original assay before subsetting. The strata are
descriptive analyses of fixed predictions and observed assays.

We computed Pearson correlation separately within each assay and bin,
requiring at least three matched measurements and variation in the labels
and in both models' predictions. We averaged the assay correlations
with weights equal to the numbers of contributing measurements. Score
uncertainty is one SE from 10,000 paired within-assay bootstrap replicates
with NumPy seed 20261112.
A dash indicates that the bin was empty or had no scorable assays.

\subsection{Ligand similarity}

We used the standardized ligands from Appendix~\ref{app:data_preprocessing}
and computed Tanimoto similarity on chirality-enabled, radius-2 binary
Morgan fingerprints with 2,048 bits~\citep{rogers2010extended,rdkit}.
For each evaluation ligand, we recorded the maximum similarity to any
retained training ligand. We also computed a separate maximum restricted
to ligands measured against the same UniProt accession. We assigned
ligands without such a reference to their own category. An assay can
contribute measurements to multiple bins. All OpenFE and FEP+4 rows
fall in the no-same-target category for the restricted analysis.

\begin{table}[!htbp]
\centering
\small
\setlength{\tabcolsep}{3pt}
\caption{Maximum ligand Tanimoto similarity within the same UniProt accession. Values are compound-weighted within-assay Pearson $r$ $\pm$ one bootstrap SE. Boltz-2 uses molecular-weight correction.}
\label{tab:similarity-same-target-scores}
\resizebox{\linewidth}{!}{%
\begin{tabular}{@{}llccccc@{}}
\toprule
Benchmark & Model & $<0.40$ & $0.40\text{--}<0.60$ & $0.60\text{--}<0.90$ & $\geq 0.90$ & No same-target ligands \\
\midrule
ChEMBL35 & STORM & $0.269 \pm 0.014$ & $0.298 \pm 0.014$ & $0.386 \pm 0.014$ & $0.146 \pm 0.108$ & $0.353 \pm 0.021$ \\
 & Boltz-2 & $0.264 \pm 0.013$ & $0.224 \pm 0.014$ & $0.331 \pm 0.015$ & $0.372 \pm 0.106$ & $0.358 \pm 0.023$ \\
OpenFE & STORM & --- & --- & --- & --- & $0.576 \pm 0.024$ \\
 & Boltz-2 & --- & --- & --- & --- & $0.570 \pm 0.024$ \\
FEP+4 & STORM & --- & --- & --- & --- & $0.705 \pm 0.053$ \\
 & Boltz-2 & --- & --- & --- & --- & $0.675 \pm 0.062$ \\
CASP16 & STORM & $0.598 \pm 0.085$ & $0.565 \pm 0.094$ & $0.799 \pm 0.088$ & --- & --- \\
 & Boltz-2 & $0.507 \pm 0.096$ & $0.506 \pm 0.085$ & $0.681 \pm 0.155$ & --- & --- \\
\bottomrule
\end{tabular}%
}
\par\medskip
\setlength{\tabcolsep}{3pt}
\caption{Maximum ligand Tanimoto similarity within the same UniProt accession. Counts are scorable assays / contributing rows.}
\label{tab:similarity-same-target-counts}
\begin{tabular}{@{}lccccc@{}}
\toprule
Benchmark & $<0.40$ & $0.40\text{--}<0.60$ & $0.60\text{--}<0.90$ & $\geq 0.90$ & No same-target ligands \\
\midrule
ChEMBL35 & 227 / 4,651 & 251 / 4,336 & 212 / 4,061 & 18 / 77 & 56 / 1,748 \\
OpenFE & 0 / 0 & 0 / 0 & 0 / 0 & 0 / 0 & 58 / 873 \\
FEP+4 & 0 / 0 & 0 / 0 & 0 / 0 & 0 / 0 & 4 / 87 \\
CASP16 & 2 / 62 & 2 / 59 & 1 / 16 & 0 / 0 & 0 / 0 \\
\bottomrule
\end{tabular}
\end{table}

\begin{table}[!htbp]
\centering
\small
\setlength{\tabcolsep}{3pt}
\caption{Maximum ligand Tanimoto similarity across the full retained training corpus. Values are compound-weighted within-assay Pearson $r$ $\pm$ one bootstrap SE. Boltz-2 uses molecular-weight correction.}
\label{tab:similarity-corpus-scores}
\resizebox{\linewidth}{!}{%
\begin{tabular}{@{}llcccc@{}}
\toprule
Benchmark & Model & $<0.40$ & $0.40\text{--}<0.60$ & $0.60\text{--}<0.90$ & $\geq 0.90$ \\
\midrule
ChEMBL35 & STORM & $0.168 \pm 0.027$ & $0.288 \pm 0.011$ & $0.348 \pm 0.011$ & $0.450 \pm 0.024$ \\
 & Boltz-2 & $0.177 \pm 0.027$ & $0.257 \pm 0.011$ & $0.283 \pm 0.013$ & $0.396 \pm 0.025$ \\
OpenFE & STORM & $0.460 \pm 0.100$ & $0.474 \pm 0.061$ & $0.599 \pm 0.038$ & $0.748 \pm 0.049$ \\
 & Boltz-2 & $0.347 \pm 0.091$ & $0.500 \pm 0.059$ & $0.599 \pm 0.040$ & $0.689 \pm 0.046$ \\
FEP+4 & STORM & --- & --- & $0.727 \pm 0.061$ & $0.837 \pm 0.093$ \\
 & Boltz-2 & --- & --- & $0.659 \pm 0.089$ & $0.881 \pm 0.050$ \\
CASP16 & STORM & $0.526 \pm 0.226$ & $0.588 \pm 0.079$ & $0.773 \pm 0.062$ & $0.671 \pm 0.253$ \\
 & Boltz-2 & $0.625 \pm 0.150$ & $0.550 \pm 0.083$ & $0.721 \pm 0.103$ & $0.625 \pm 0.220$ \\
\bottomrule
\end{tabular}%
}
\par\medskip
\setlength{\tabcolsep}{3pt}
\caption{Maximum ligand Tanimoto similarity across the full retained training corpus. Counts are scorable assays / contributing rows.}
\label{tab:similarity-corpus-counts}
\begin{tabular}{@{}lcccc@{}}
\toprule
Benchmark & $<0.40$ & $0.40\text{--}<0.60$ & $0.60\text{--}<0.90$ & $\geq 0.90$ \\
\midrule
ChEMBL35 & 87 / 1,080 & 346 / 6,621 & 296 / 5,588 & 95 / 1,400 \\
OpenFE & 10 / 102 & 21 / 210 & 32 / 334 & 12 / 157 \\
FEP+4 & 0 / 0 & 0 / 0 & 3 / 59 & 3 / 25 \\
CASP16 & 2 / 17 & 2 / 74 & 2 / 38 & 1 / 11 \\
\bottomrule
\end{tabular}
\end{table}

\subsection{Training protein sequence identity}

We binned each evaluation protein by aligned identity to its
highest-bitscore MMseqs2 hit in the retained training targets.
Exact sequence matches were assigned 100\% identity. For evaluation
constructs that differ from the training sequence, the bins use aligned
identity even when the accession is shared. Identity refers to the
aligned region. On OpenFE, median query and training-hit coverage were
98.9\% and 97.7\%, with minima of 10.5\% and 2.3\%. The corresponding
FEP+4 minima were 97.5\% and 96.4\%.

OpenFE has 18 of 39 targets below 60\% aligned identity. ChEMBL35 has one
target in the below-30\% bin and 246 of 276 matched targets in the
at-least-90\% bin. All FEP+4 targets fall between 70\% and 90\%,
and both CASP16 targets are in the at-least-90\% bin.

\begin{table}[!htbp]
\centering
\small
\setlength{\tabcolsep}{3pt}
\caption{Aligned identity to the highest-bitscore retained training protein (percent). Values are compound-weighted within-assay Pearson $r$ $\pm$ one bootstrap SE. Boltz-2 uses molecular-weight correction.}
\label{tab:similarity-identity-scores}
\resizebox{\linewidth}{!}{%
\begin{tabular}{@{}llcccccccc@{}}
\toprule
Benchmark & Model & $<30$ & $30\text{--}<40$ & $40\text{--}<50$ & $50\text{--}<60$ & $60\text{--}<70$ & $70\text{--}<80$ & $80\text{--}<90$ & $\geq 90$ \\
\midrule
ChEMBL35 & STORM & $0.862 \pm 0.085$ & $0.594 \pm 0.057$ & $0.381 \pm 0.057$ & $0.466 \pm 0.043$ & $0.314 \pm 0.054$ & $0.232 \pm 0.050$ & $0.195 \pm 0.052$ & $0.355 \pm 0.008$ \\
 & Boltz-2 & $0.720 \pm 0.156$ & $0.634 \pm 0.052$ & $0.199 \pm 0.059$ & $0.540 \pm 0.044$ & $0.398 \pm 0.057$ & $0.245 \pm 0.069$ & $0.197 \pm 0.050$ & $0.296 \pm 0.008$ \\
OpenFE & STORM & $0.636 \pm 0.164$ & $0.651 \pm 0.153$ & $0.422 \pm 0.050$ & $0.538 \pm 0.069$ & $0.579 \pm 0.124$ & $0.612 \pm 0.043$ & $0.697 \pm 0.046$ & --- \\
 & Boltz-2 & $0.338 \pm 0.169$ & $0.906 \pm 0.050$ & $0.320 \pm 0.055$ & $0.613 \pm 0.068$ & $0.410 \pm 0.140$ & $0.615 \pm 0.042$ & $0.764 \pm 0.042$ & --- \\
FEP+4 & STORM & --- & --- & --- & --- & --- & $0.721 \pm 0.059$ & $0.653 \pm 0.120$ & --- \\
 & Boltz-2 & --- & --- & --- & --- & --- & $0.634 \pm 0.080$ & $0.805 \pm 0.061$ & --- \\
CASP16 & STORM & --- & --- & --- & --- & --- & --- & --- & $0.671 \pm 0.045$ \\
 & Boltz-2 & --- & --- & --- & --- & --- & --- & --- & $0.640 \pm 0.052$ \\
\bottomrule
\end{tabular}%
}
\par\medskip
\setlength{\tabcolsep}{3pt}
\caption{Aligned identity to the highest-bitscore retained training protein (percent). Counts are distinct targets / scorable assays / contributing rows.}
\label{tab:similarity-identity-counts}
\resizebox{\linewidth}{!}{%
\begin{tabular}{@{}lcccccccc@{}}
\toprule
Benchmark & $<30$ & $30\text{--}<40$ & $40\text{--}<50$ & $50\text{--}<60$ & $60\text{--}<70$ & $70\text{--}<80$ & $80\text{--}<90$ & $\geq 90$ \\
\midrule
ChEMBL35 & 1 / 1 / 24 & 3 / 3 / 151 & 5 / 9 / 260 & 5 / 13 / 419 & 3 / 9 / 220 & 5 / 10 / 298 & 8 / 11 / 376 & 246 / 387 / 13393 \\
OpenFE & 3 / 3 / 21 & 2 / 2 / 24 & 6 / 11 / 213 & 7 / 8 / 108 & 2 / 2 / 37 & 11 / 16 / 270 & 8 / 16 / 200 & 0 / 0 / 0 \\
FEP+4 & 0 / 0 / 0 & 0 / 0 / 0 & 0 / 0 / 0 & 0 / 0 / 0 & 0 / 0 / 0 & 3 / 3 / 66 & 1 / 1 / 21 & 0 / 0 / 0 \\
CASP16 & 0 / 0 / 0 & 0 / 0 / 0 & 0 / 0 / 0 & 0 / 0 / 0 & 0 / 0 / 0 & 0 / 0 / 0 & 0 / 0 / 0 & 2 / 2 / 140 \\
\bottomrule
\end{tabular}%
}
\end{table}

\end{document}